\documentclass[twocolumn,secnumarabic,amssymb,nobibnotes,aps,nofootinbib,noeprint]{revtex4-1}
\usepackage[english]{babel}
\usepackage{graphicx}

\usepackage{array}
\usepackage{amsmath}
\usepackage{amsfonts}
\usepackage[usenames,dvipsnames]{color}
\usepackage{hyperref}
\usepackage{rotating}
\usepackage{booktabs} % To set batter tables, see: booktabs docu
\usepackage{slashed}
\usepackage{transparent}

\usepackage{ulem}

\DeclareMathAlphabet{\mathbb}{U}{bbold}{m}{n}

\begin{document}
\title{Three-body correlations and conditional forces in suspensions of active hard disks}
\author{Andreas H\"{a}rtel,$^{1}$ David Richard,$^2$ and Thomas Speck$^2$} 
\affiliation{$^1$ Institute of Physics, University of Freiburg, 
	Hermann-Herder-Stra\ss{}e 3, 79104 Freiburg, Germany\\
$^2$ Institute of Physics, 
	Johannes Gutenberg-University Mainz, 
	Staudinger Weg 9, 55128 Mainz, Germany\\
	-- published in: Physical Review E \textbf{97}, 012606 (2018), \href{http://dx.doi.org/10.1103/PhysRevE.97.012606}{DOI: 10.1103/PhysRevE.97.012606} -- 
}
%\date{\today}

\begin{abstract}
Self-propelled Brownian particles show rich out-of-equilibrium 
physics, for instance, the motility-induced phase separation (MIPS). 
While decades of studying the structure of liquids have established a deep understanding 
of passive systems, not much is known about correlations in active suspensions. 
In this work we derive an approximate analytic theory for three-body correlations and forces in 
systems of active Brownian disks starting from the many-body Smoluchowski equation. 
We use our theory to predict the conditional forces that act on a tagged particle and their 
dependence on the propulsion speed of self-propelled disks. We identify preferred 
directions of these forces in relation to the direction of propulsion and the positions of 
the surrounding particles. We further relate our theory to the effective swimming 
speed of the active disks, which is relevant for the physics of MIPS. 
To test and validate our theory, we additionally run particle-resolved computer simulations, 
for which we explicitly calculate the three-body forces. 
In this context, we discuss the modeling of active Brownian swimmers with nearly hard 
interaction potentials. We find very good agreement between our simulations and numerical 
solutions of our theory, especially for the nonequilibrium pair-distribution function. 
For our analytical results, we carefully discuss their range of validity 
in the context of the different levels of approximation we applied. 
This discussion allows us to study the individual contribution of particles to three-body forces 
and to the emerging structure. Thus, our work sheds light on the 
collective behavior, provides the basis for further studies of correlations in active 
suspensions, sheds new light onto the collective behavior, and 
makes a step towards an emerging liquid state theory. 
\end{abstract}

\maketitle

\section{Introduction}

Research on active matter recently revealed exciting new phenomena 
at the intersection of 
physics, chemistry, and biology \cite{
barry_pnas107_2010,% Dictyostelium amoebae and neutrophils can swim
ramaswamy_arcmp1_2010,% Review: The Mechanics and Statistics of Active Matter
schaller_nature467_2010,% Polar patterns of driven filaments
liu_science334_2011,% Sequential Establishment of Stripe Patterns in an Expanding Cell Population (Escherichia coli cells)
sanchez_nature491_2012,% Spontaneous motion in hierarchically assembled active matter
palacci_science339_2013,% Living Crystals of Light-Activated Colloidal Surfers
marchetti_rmp85_2013,% Hydrodynamics of soft active matter
bricard_nature503_2013,% Emergence of macroscopic directed motion in populations of motile colloids
cates_arcmp6_2015,% Review on ``Motility-Induced Phase Separation''
goldstein_arfm47_2015,% Model organisms for active swimmers are e.g. green algae
wang_acr48_2015,% From One to Many: Dynamic Assembly and Collective Behavior of Self-Propelled Colloidal Motors
yan_nm15_2016,% Reconfiguring active particles by electrostatic imbalance
niu_prl119_2017% Self-Assembly of Colloidal Molecules due to Self-Generated Flow
}.
It deals with 
particles and individuals that show self-propelled motion, which 
includes living ``matter'' like fish, flocks of birds \cite{cagna_pnas107_2010}, 
and bacteria \cite{liu_science334_2011,wensink_pnas109_2012}, as well as 
artificial colloidal swimmers \cite{howse_prl99_2007,gangwal_prl100_2008,palacci_prl105_2010,
theurkauff_prl108_2012,palacci_science339_2013,buttinoni_prl110_2013,wang_acr48_2015,niu_prl119_2017} 
and robots \cite{rubenstein_science345_2014}. 
Accordingly, detailed knowledge of the fundamental mechanisms that drive active systems is important 
to understand and control swimming mechanisms and self-organization phenomena such as 
collective motion \cite{marchetti_rmp85_2013,wysocki_epl105_2014}, 
phase separation due to motility differences \cite{cates_arcmp6_2015,speck_epjst225_2016}, 
and formation of periodic stripe patterns \cite{liu_science334_2011}. 
The rich variation of nonequilibrium 
phenomena in active matter results in potential applications in self-assembly and materials 
research \cite{vandermeer_sm12_2016}. 

The fundamental mechanisms in active many-body systems can be studied with methods from 
out-of-equilibrium statistical physics. 
Beyond the well-studied behavior of equilibrated passive systems, new concepts are 
needed in active systems, for instance, 
to define pressure \cite{solon_prl114_2015,speck_pre93_2016}. 
The motion of active particles is governed by many different driving mechanisms such as 
amoeboid or human swimming \cite{barry_pnas107_2010,farutin_prl111_2013}, 
running of animals on land \cite{toner_pre58_1998}, 
phoretic motion \cite{anderson_ar21_1989,gangwal_prl100_2008,palacci_science339_2013}, 
use of flagella \cite{reichert_epje17_2005,yang_pre89_2014}, 
and rocket propulsion where fuel is expelled. 
Depending on whether their shapes and pair interactions are apolar or polar 
\cite{ramaswamy_arcmp1_2010,marchetti_rmp85_2013}, active particles can also show nematic ordering 
\cite{marenduzzo_pre76_2007,cates_prl101_2008,ramaswamy_arcmp1_2010,schaller_nature467_2010,
wensink_pnas109_2012,marchetti_rmp85_2013}. 
Further, the coupling of active particles to hydrodynamic interactions determines whether 
systems behave wet or dry, where the theoretical description of dry systems does not include an 
explicit solvent \cite{marchetti_rmp85_2013}. 
For this reason, 
the identification of model organisms \cite{goldstein_arfm47_2015} 
and minimal models \cite{
vicsek_prl75_1995,% Novel Type of Phase Transition in a System of Self-Driven Particles
czirok_pa281_2000,% Collective behavior of interacting self-propelled particles
tailleur_prl100_2008,% original work on MIPS
nash_prl104_2010,% Run-and-Tumble Particles with Hydrodynamics: Sedimentation, Trapping, and Upstream Swimming
cates_rpp75_2012,
cates_epl101_2013,% When are active Brownian particles and run-and-tumble particles equivalent? Consequences for motility-induced phase separation
bialke_epl103_2013,% 2-body level
farutin_prl111_2013,% Amoeboid Swimming: A Generic Self-Propulsion of Cells in Fluids by Means of Membrane Deformations
das_prl112_2014,% Phase Behavior of Active Swimmers in Depletants: Molecular Dynamics and Integral Equation Theory
nagai_prl114_2015,% Collective motion of self-propelled particles with memory
ni_prl114_2015,% Tunable Long Range Forces Mediated by Self-Propelled Colloidal Hard Spheres
wysocki_pre91_2015,% Also shape asymmetries have been studied and their effect on the adsorption capacity at confining channel walls
menzel_jcp144_2016% Dynamical density functional theory for microswimmers
} is important to isolate and study basic principles. 

One minimal model for active matter is the model of active Brownian particles, which 
combines volume exclusion and Brownian directed motion but neglects long-range phoretic and 
hydrodynamic interactions. 
Accordingly, this model of ``scalar active matter'' solely involves scalar fields 
\cite{tiribocchi_prl115_2015}. 
The model shows many phenomena when self-propelled individuals (swimmers) interact with 
surfaces, channels, and traps \cite{
nash_prl104_2010,% Run-and-Tumble Particles with Hydrodynamics: Sedimentation, Trapping, and Upstream Swimming
wysocki_pre91_2015,%shape asymmetries and their effect on the adsorption capacity at confining channel walls
menzel_jcp144_2016% Traped in a harmonic potential, the swimmers concentrate at the outer border of the trap for 
}
or with additional passive particles \cite{das_prl112_2014,wittkowski_njp19_2017}. 
In bulk it describes a motility-induced phase separation (MIPS)
\cite{tailleur_prl100_2008,buttinoni_prl110_2013,bialke_epl103_2013,speck_prl112_2014,cates_arcmp6_2015}, 
where repulsive Brownian swimmers separate in dense and dilute phases at sufficiently high propulsion speeds 
and number densities even in the absence of cohesive forces. 

To unveil the fundamental mechanism of MIPS, previous and the present work 
use the Smoluchowski equation \cite{hansen_book_2013} 
for the time evolution of the distribution of particle positions 
\cite{bialke_epl103_2013,speck_jcp142_2015,wittkowski_njp19_2017}. 
Until now, the set of hierarchically connected equations was 
closed only on the two-particle level \cite{bialke_epl103_2013,wittkowski_njp19_2017}, 
which already allows to define an anisotropy parameter $\zeta_1$ that describes the anisotropy of the 
pair-distribution function around a tagged particle \cite{bialke_epl103_2013}. 
The parameter $\zeta_1$ is strongly correlated to the propulsion speed of a 
single particle and presents a key ingredient for the theoretical description of MIPS \cite{bialke_epl103_2013}. 
To go beyond one-body densities and in order to {\it a priori} predict two-body correlations, forces, and 
effective swimming speeds, one has to consider three-body correlations. 
This is the aim of the present work. 
Studying them and finding reasonable approximations will allow us to set up an analytical theory 
that describes conditional three-body forces and their preferred directions for self-propelled Brownian 
particles. Moreover, it will enable us to define effective hard-disk coefficients that have the potential 
to act as order parameters for active systems.  

Already in passive colloidal systems not much work has explicitly addressed three-body correlations 
\cite{zahn_prl91_2003,bartnick_jpcm28_2016} and three-body forces actually have 
not explicitly been reported in this field at all. One reason might be the difficulty of finding an 
adequate closure on the 
three-body level \cite{kirkwood_jcp3_1935,deboer_physica6_1939,stillinger_jcp57_1972,fischer_mp33_1977,
lee_jcp135_2011}. One common closure is the superposition approximation by Kirkwood 
\cite{kirkwood_jcp3_1935,vanKampen_physica27_1961,zahn_prl91_2003}, which 
shows reasonable structural agreement with simulations \cite{zahn_prl91_2003} even 
if it is just a first-order expansion of the triplet distribution function \cite{barker_rmp48_1976}. 
Thus, research beyond the typical study of two-body correlations might give additional insight 
into correlations and structure even in passive systems. 

In the present work we study three-body correlations and forces in suspensions of active Brownian particles 
using theory and simulations. In Sec.~II we develop the general theoretical framework 
beyond the two-body level based on the Smoluchowski equation for active Brownian particles. 
In order to find a closed form of our theory, we apply the 
Kirkwood superposition approximation. We further focus on the special case of completely 
steric pair interactions (hard disks) to achieve analytical results for averaged three-body forces 
in active systems. In Sec.~III we first present data from Brownian dynamics 
simulations. Then we compare these data with our analytical results. In addition, we 
solve our theoretical framework numerically. By comparing our results from these 
numerical calculations, the analytical theory, and the simulations, we establish the range 
of validity and identify limitations of our theory. 
We discuss in Sec.~IV our results and theoretical predictions for active systems and 
summarize in Sec.~V.

\section{Theory}

\label{sec:theory}

In this section we derive step by step an analytical theory for the microscopic structure of 
active Brownian particles that interact via a pair potential. Intermediate results are 
valid for general pair interactions and some of these results are even exact. We structure our derivation 
as follows. First, in Sec.~\ref{sec:abps} we formulate the general model and framework and 
in Sec.~\ref{sec:many-body-hierarchy} we introduce the relevant variables. 
Then in Sec.~\ref{sec:symmetries-parametrization} we take advantage of symmetries 
to further reduce the number of parameters and 
in Sec.~\ref{sec:closure-twobody-level} we discuss the closure of the ensuing hierarchy of equations. 
Only then do we restrict our theory in Sec.~\ref{sec:special-case-hard-disks} to the special case of 
hard disks and simplify in Sec.~\ref{sec:simplification-hard-disks} our closure relation 
from Sec.~\ref{sec:closure-twobody-level}. Finally, in Sec.~\ref{sec:expansion-g} we expand 
the pair-distribution function to achieve our final analytic results.

\subsection{Active Brownian Particles (ABP)}
\label{sec:abps}

%+++++++++++++++++++++++++++++++++++++++
\begin{figure}
	\centering
  \includegraphics[width=8.5cm]{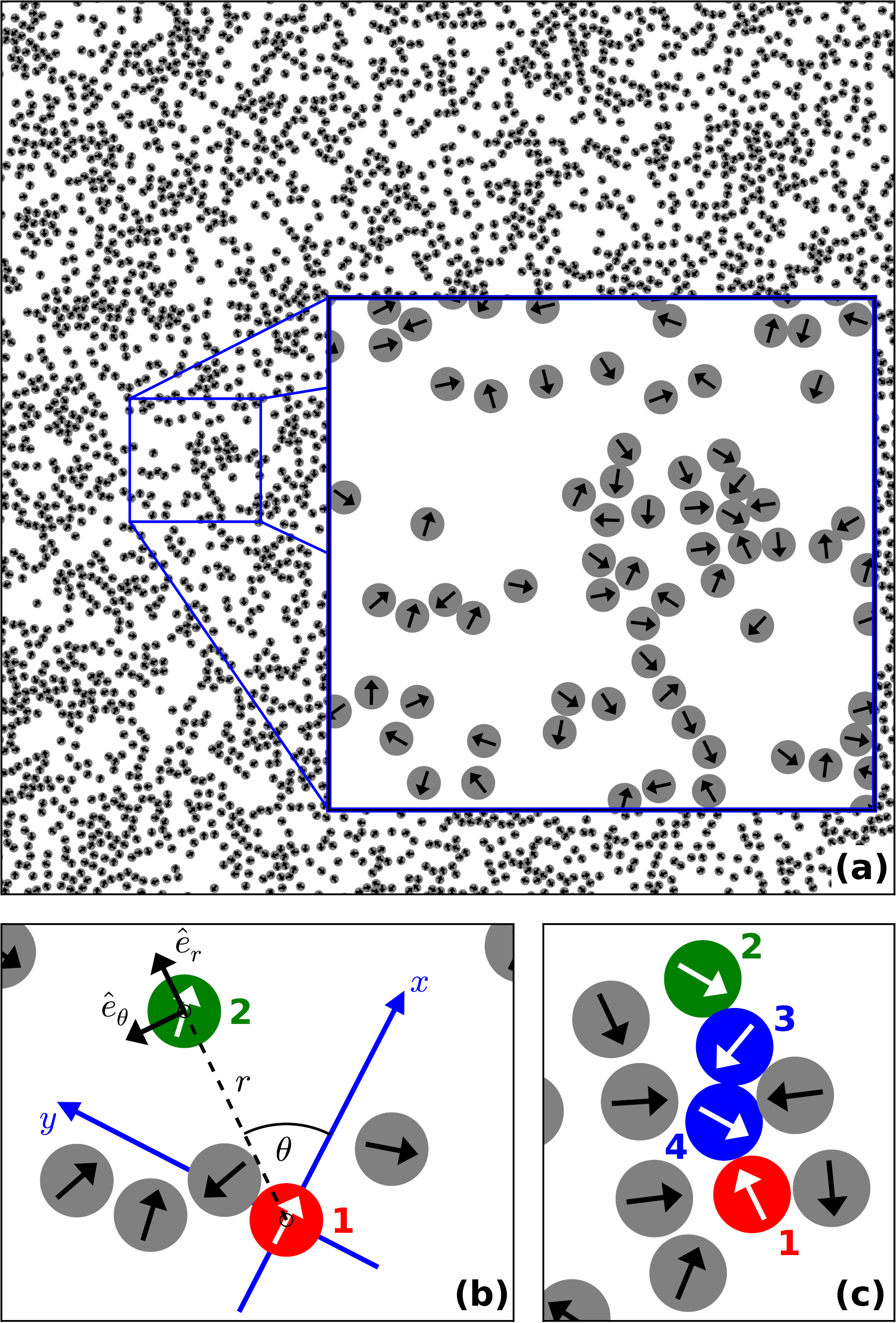}
	\caption{\label{fig:sketch-coordinates}
		Simulation snapshot of 4096 self-propelled disks at number density $\bar{\rho}=0.3$ and constant 
		propulsion speed $v_0/d_{\rm eff}=5$. The system size is $L\times L$, with $L\approx 116.85$, 
		and the directions of propulsion $\hat{e}_i$ for each particle $i$ are shown by arrows. 
		(b) Situation from within the snapshot in (a) with two tagged particles 
		and our corresponding relative coordinates. The origin is fixed at the 
		position of the first particle with the $x$ direction along its direction of propulsion. The second particle 
		is located at the position $\vec{r}=(r\cos\theta,r\sin\theta)$. 
		The normalized basis vectors $\hat{e}_r$ and $\hat{e}_\theta$ are shown for the position of particle $2$. 
		(c) Situation from within the snapshot in (a) with 
		two tagged particles interacting via two intermediate particles $3$ and $4$. 
	}
\end{figure}
%+++++++++++++++++++++++++++++++++++++++

Active Brownian particles (ABPs) are 
a minimal model of particles moving in contact with a heat bath and combining directed motion 
with volume exclusion. Although strictly speaking this model falls into the class of dry active matter 
without an explicit solvent \cite{marchetti_rmp85_2013}, 
we will use the term ``swimming'' to describe the directed motion of particles. 
We consider $N$ particles in a two-dimensional system of area $V$ with mean 
number density $\bar{\rho}=N/V$, as shown in Fig.~\ref{fig:sketch-coordinates}. 
The particles at positions $\vec{r}_k$ interact via general radial-symmetric pair 
potentials $u(r)$ with total potential energy 
\begin{align}
	U &= \sum_{k<k'} u(|\vec{r}_k-\vec{r}_{k'}|) . \label{eq:total-potential-energy}
\end{align}
Every particle is self-propelled, i.e., it swims with a constant propulsion 
speed $v_0$ in the direction 
\begin{align}
\hat{e}_k &= \begin{pmatrix} \cos(\varphi_k) \\ \sin(\varphi_k) \end{pmatrix} . 
\end{align}

The coupled equations of motion for the particle positions $\vec{r}_k$ and orientations $\hat{e}_k$ are 
\begin{align}
	\dot{\vec{r}}_k &= -\mu_0\vec{\nabla}_k U + v_0\hat{e}_k + \vec{\xi}_k , \label{eq:eom} \\
	\dot{\hat{e}}_k &= \vec{\eta}_k \times \hat{e}_k , 
\end{align}
with a mobility $\mu_0$ and the white Gaussian noises $\vec{\xi}_k$ and $\vec{\eta}_k$. 
We write $\vec{\nabla}_k U(\vec{r}_1,\dots,\vec{r}_N)$ for the partial gradient of the scalar 
function $U$, where the gradient is only taken with respect to the $k$-th parameter $\vec{r}_k$ while 
all remaining $\vec{r}_i$ with $i\neq k$ are kept fixed. 
The white Gaussian noises have zero mean and temporal mean-square deviations 
\begin{align}
	\langle \vec{\xi}_k(t) \otimes \vec{\xi}_{k'}(t') \rangle &= 2 D_0 \Bbb{1} \delta_{kk'}\delta(t-t') , \\
	\langle \vec{\eta}_k(t) \otimes \vec{\eta}_{k'}(t') \rangle &= 2 D_{\rm r} \Bbb{1} \delta_{kk'}\delta(t-t') . 
\end{align}
Here $\Bbb{1}$ denotes the identity matrix. 
We assume that the spatial diffusion constant $D_0$ and the rotational diffusion constant $D_{\rm r}$ 
are hydrodynamically coupled by $D_{\rm r}=3 D_0/\sigma^2$ \cite{einstein_apb19_1906} 
such that the no-slip boundary condition holds as in previous work \cite{bialke_epl103_2013,richard_sm12_2016}; 
$\sigma$ is the (effective) particle diameter. 

Throughout this work we employ dimensionless quantities and measure lengths in units of $\sigma$, 
time in units of $\sigma^2/D_0$, and energy in units of $k_{\rm B}T$. Here $k_{\rm B}$ denotes 
Boltzmann's constant and $T$ is the temperature of the system. 
Consequently, we use $D_{\rm r}=3$.

\subsection{Many-body hierarchy}
\label{sec:many-body-hierarchy}

The time evolution of the 
probability density $P_N(\vec{r}\,^{(N)},\varphi^{(N)};t)$ to find $N$ particles at positions 
$\vec{r}\,^{(N)}$ with directions of propulsion (orientations) denoted by the 
angles $\varphi^{(N)}$ is governed by the Smoluchowski equation \cite{hansen_book_2013} 
\begin{align}
	\partial_t P_N = & 
	\sum_{k=1}^N \vec{\nabla}_k \cdot \left[ (\vec{\nabla}_k U) 
	- v_0\hat{e}_k + \vec{\nabla}_k \right] P_N \notag \\
	&
	+ D_{\rm r}\sum_{k=1}^N \partial_{\varphi_k}^2 P_N \label{eq:smoluchowski} . 
\end{align}
We use $\vec{r}\,^{(n)}$ as a multi-index denotation for $(\vec{r}_1,\dots,\vec{r}_n)$. 
The joint probability distribution $P_N$ is normalized to unity, i.e., $\int \dots \int P_N = 1$.  
Then we define a hierarchy of $n$-body densities 
$\Psi_n\equiv\Psi_n(\vec{r}\,^{(n)},\varphi^{(n)};t)$ for $1\leq n\leq N$ by 
\begin{align}
	& \Psi_n(\vec{r}\,^{(n)},\varphi^{(n)};t) \notag \\
	& \quad\quad = \int d\vec{r}_{n+1} \dots d\vec{r}_N 
	\int d\varphi_{n+1} \dots d\varphi_N \frac{N!}{(N-n)!} P_N . 
	\label{eq:density-projection-operator}
\end{align}
The $n$-body number densities $\rho_n = \int d\varphi^{(n)} \Psi_{n}$ at a certain time $t$ are achieved 
by integrating out the orientations. We further define a conditional one-body probability $P_1$ in order to 
describe $\Psi_3$ in terms of $\Psi_2$, i.e., 
\begin{align}
	\Psi_3(\vec{r}\,^{(3)},\varphi^{(3)};t) =& 
	\Psi_2(\vec{r}\,^{(2)},\varphi^{(2)};t) 
	\frac{N-2}{V} \notag \\
	& \times 
	V P_1(\vec{r}_3,\varphi_3|\vec{r}\,^{(2)},\varphi^{(2)};t) . 
	\label{eq:conditional-probability2}
\end{align}
We also define the conditional distribution 
\begin{align}
	g_1(\vec{r}_3|\vec{r}\,^{(2)},\varphi^{(2)};t) &= 
	V \int_0^{2\pi} d\varphi_3 \, P_1(\vec{r}_3,\varphi_3|\vec{r}\,^{(2)},\varphi^{(2)};t) , 
\end{align}
which describes the distribution of a (third) particle when two particles $1$ and $2$ are 
given with positions $\vec{r}\,^{(2)}$ and orientational angles $\varphi^{(2)}$. 
Note that in the limit of large $N$ the factor $(N-2)/V\to\bar{\rho}$. 

The integration $\int d\vec{r}_3 \dots d\vec{r}_N\int d\varphi_3 \dots d\varphi_N (N-1)N$ 
on both sides of the Smoluchowski equation (\ref{eq:smoluchowski}) leads to 
\begin{align}
	& \partial_t \Psi_2(\vec{r}_1,\varphi_1,\vec{r}_2,\varphi_2;t)
	= \sum_{k=1,2} \bigg( \notag \\
	%%%
	&\quad 
	- \vec{\nabla}_k \cdot \left[ -\big(\vec{\nabla}_k u(|\vec{r}_1-\vec{r}_2|)\big) 
	+ \vec{F}_k + v_0\hat{e}_k - \vec{\nabla}_k \right] \notag \\
	%%%
	&\quad\quad\quad\quad\quad \times \Psi_2(\vec{r}_1,\varphi_1,\vec{r}_2,\varphi_2;t) \notag \\
	%%%
	&\quad %\quad\quad\quad\quad\quad 
	+ D_{\rm r}\partial_{\varphi_k}^2 \Psi_2(\vec{r}_1,\varphi_1,\vec{r}_2,\varphi_2;t) \bigg) \label{eq:eom2} , 
\end{align}
with the conditional forces 
\begin{align}
	& \vec{F}_k(\vec{r}_1,\varphi_1,\vec{r}_2,\varphi_2;t) = \notag \\
	& -\bar{\rho} \int d\vec{r}_3 u'\big(|\vec{r}_k-\vec{r}_3|\big)\frac{\vec{r}_k-\vec{r}_3}{|\vec{r}_k-\vec{r}_3|} 
	g_1(\vec{r}_3|\vec{r}_1,\varphi_1,\vec{r}_2,\varphi_2;t) . 
	\label{eq:force-3particle}
\end{align}
These terms describe the summed contribution of all forces 
$\vec{F}_{i\to k}$ acting from a particle $i\in\{3,\dots,N\}$ on the respective particle 
$k\in\{1,2\}$ in the presence of the remaining second particle, 
i.e., $\vec{F}_k = \sum_{i=3}^{N}\vec{F}_{i\to k}$. 
This is illustrated in Fig.~\ref{fig:sketch-coordinates}(b), where particles $1$ and $2$ are 
shown in red and green, respectively. All third particles 
that contribute to the conditional forces $\vec{F}_{k}$ are shown in gray. 
Note that this formalism is not restricted to a specific pair interaction between 
individual particles; we only assumed rotational symmetry. 
In the special case of hard interactions most of the contributions 
of the gray particles in Fig.~\ref{fig:sketch-coordinates}(b) would vanish, 
because hard pair interactions lead to forces only when particles touch. For example, 
only one gray particle in Fig.~\ref{fig:sketch-coordinates}(b) would contribute a 
nonvanishing force. Similarly, only the blue particles with indices $3$ and $4$ 
in Fig.~\ref{fig:sketch-coordinates}(c) 
would contribute to the direct forces $\vec{F}_1$ and $\vec{F}_2$ in this case. 
Later we will see that particle 1 is influenced by the presence of particle 2 in the situation 
shown in Fig.~\ref{fig:sketch-coordinates}(c) via both blue particles 3 and 4. 
A consequence of the rare event of a particle contact in the case of hard disks is that 
statistical averaging must be performed over a much larger number of snapshots than in the case 
of softer interactions, at least when aiming at a similar quality of statistics.

\subsection{Symmetries and parametrization}

\label{sec:symmetries-parametrization}

In the following we focus on the homogeneous phase such that the two-body density 
$\Psi_2(\vec{r}\,^{(2)},\varphi^{(2)};t)$ depends only on the displacement vector $\vec{r}_2-\vec{r}_1$. 
Note that this assumption does not rule out the ability of our theory to study phase separations 
like MIPS, because the theory is still able to describe both phases individually. 
Further, divergent behavior in a theory for a homogeneous phase may indicate phase 
instabilities and thus be a signature of other phases. 
Our theory still holds for any rotationally symmetric pair-interaction potential $u(r)$. 
Provided the assumption of a homogeneous phase, we change 
to relative coordinates in the reference frame of a tagged particle, say, 
particle 1, such that the tagged particle is oriented in the $x$ direction and its position 
$\vec{r}_1$ becomes the origin of our coordinate system. 
Accordingly, the set $\{\vec{r}_1,\varphi_1,\vec{r}_2,\varphi_2\}$ of parameters reduces 
to the relative position and orientation of the second particle with respect to the first 
one, as sketched in Fig.~\ref{fig:sketch-coordinates}(b). 
We parametrize the relative position by $\vec{r}=(r \cos\theta , r \sin\theta)$ 
such that the normalized directions of the circular coordinates $r$ and $\theta$ 
are $\hat{e}_r=(\cos\theta,\sin\theta)$ and $\hat{e}_{\theta}=(-\sin\theta,\cos\theta)$. 
For completeness, the gradient and divergence operators 
for a vector $\vec{A}$ and a scalar $A$ in these polar coordinates read 
\begin{align}
\vec{\nabla} \cdot \vec{A} &= \frac{1}{r} \frac{\partial}{\partial r} \Big( r \hat{e}_r\cdot\vec{A} \Big)
	+ \frac{1}{r} \frac{\partial}{\partial\theta} \Big( \hat{e}_{\theta} \cdot \vec{A} \Big) , 
	\label{eq:gradient-operator} \\
\vec{\nabla} A &= \frac{\partial A}{\partial r} \hat{e}_r	
	+ \frac{1}{r} \frac{\partial A}{\partial\theta} \hat{e}_{\theta} . 
	\label{eq:divergence-operator}
\end{align}
We further transform the two-body density from Eq.~(\ref{eq:eom2}) into the form of 
a pair-distribution function by 
integrating out the orientation $\varphi_2$ of the second particle 
and multiplying by a factor $2\pi/\bar{\rho}^2$, where again we use $(N-1)/V\to\bar{\rho}$ 
for large $N$. Accordingly, we obtain 
\begin{align}
	\frac{2\pi}{\bar{\rho}} \frac{V}{N-1} \int_{0}^{2\pi}d\varphi_2 
	\Psi_2(\vec{r}_1,\varphi_1,\vec{r}_2,\varphi_2;t) 
\stackrel{{\tiny{\begin{matrix}\vec{r}_2-\vec{r}_1=\vec{r}\\\varphi_1=0\end{matrix}}}}{\to}
	g(r,\theta;t) \label{eq:definition-g}
\end{align}
and the Smoluchowski equation (\ref{eq:eom2}) becomes 
\begin{align}
\partial_t g(r,\theta;t) 
&= \vec{\nabla}\cdot\bigg[ 
  - 2\Big(\vec{\nabla} u(r)\Big) 
	+ \vec{F}_1\big(r,\theta;t\big)
- \vec{F}_2\big(r,\theta;t\big) \notag \\
&\quad\quad + v_0\hat{e}_1 + 2 \vec{\nabla} 
\bigg] g(r,\theta;t) 
+ D_{\rm r} \partial_{\theta}^2 g(r,\theta;t)  
\label{eq:eom5b} 
\end{align}
for the pair-distribution function $g(r,\theta;t)$. 
Consequently, the conditional forces from Eq.~(\ref{eq:force-3particle}) now read 
\begin{align}
	& \vec{F}_1(\vec{r};t) = \notag \\
	&\quad\quad -\bar{\rho} \int d\varphi_2 \int d\vec{r}\,' \, u'\big(|\vec{r}\,'\,|\big) 
	\frac{-\vec{r}\,'}{|\vec{r}\,'\,|} 
	g_1\Big(\vec{r}\,'\,\big|\,0,0,\vec{r},\varphi_2;t\Big) \label{eq:force1b} , \\
	%%%
	& \vec{F}_2(\vec{r};t) = \notag \\
	&\quad\quad -\bar{\rho} \int d\varphi_2 \int d\vec{r}\,' \, u'\big(|\vec{r}\,'\,|\big) 
	\frac{-\vec{r}\,'}{|\vec{r}\,'\,|} 
	g_1\Big(\vec{r}\,'\,\big|-\vec{r},0,0,\varphi_2;t\Big) \label{eq:force2b} . 
\end{align}

\subsection{Closure on the two-body level}

\label{sec:closure-twobody-level}

In order to obtain a closed form of Eq.~(\ref{eq:eom5b}), we 
have to determine the conditional distribution $g_1(\vec{r}_3|\dots)$ that enters the force terms 
from Eqs.~(\ref{eq:force1b}) and (\ref{eq:force2b}). 
For this purpose, we apply the Kirkwood superposition approximation 
\cite{kirkwood_jcp3_1935,vanKampen_physica27_1961,zahn_prl91_2003}, 
which is attained by the first order of a diagrammatic expansion of the triplet distribution 
function \cite{barker_rmp48_1976}, i.e., 
\begin{align}
	g_{123} &= g_{12} g_{13} g_{23} \left[ 1 + \int d\vec{r}_4 \int d\varphi_4 f_{14}f_{24}f_{34} + \dots \right] , 
	\label{eq:barker-henderson-correction}
\end{align}
with $f_{ij}$ the Mayer function and subscripts indicating particle indices \cite{hansen_book_2013}. 
The expansion describes the three-body distribution $g_{123}$ as the sum of products of 
pairwise distributions between (i) the three particles $1$, $2$, and $3$, (ii) the three particles 
and one additional fourth particle, (iii) five particles, and so on. The Kirkwood approximation has 
mainly been applied to systems in equilibrium, but there is no restriction apart from assuming pairwise 
particle interactions as we have introduced in Eq.~(\ref{eq:total-potential-energy}). 
By applying the Kirkwood approximation $g_{123}=g_{12}g_{13}g_{23}$ as a closure for our 
theoretical framework, we find 
\begin{align}
	&	\Psi_3(\vec{r}_1,\varphi_1,\vec{r}_2,\varphi_2,\vec{r}_3,\varphi_3;t) = 
	%%%
\Psi_2(\vec{r}_1,\varphi_1,\vec{r}_2,\varphi_2;t) \notag \\
&\quad \times g_2(\vec{r}_2,\varphi_2,\vec{r}_3,\varphi_3;t) 
g_2(\vec{r}_3,\varphi_3,\vec{r}_1,\varphi_1;t) \Psi_1(\vec{r}_3,\varphi_3;t) . \label{eq:kirkwood-gs}
\end{align}
Note that normalization is not contained within the Kirkwood approximation and that 
the equality in Eq.~(\ref{eq:kirkwood-gs}) only holds in the limit of large particle numbers, 
where $N(N-2)/(N-1)^2\approx 1$. However, our calculations are still valid for any kind of 
pair interaction $u(r)$. According to Eq.~(\ref{eq:kirkwood-gs}), we find closed terms 
for the conditional distributions that occur in the force terms from Eqs.~(\ref{eq:force1b}) 
and (\ref{eq:force2b}), i.e., 
\begin{align}
& \int d\varphi_2 \, g_1\Big( \vec{r}\,' \big| 0, 0, \vec{r}, \varphi_2; t \Big) \notag \\
&\quad\quad = \Big\langle g\big(|\vec{r}\,'-\vec{r}\,|,\varphi_2;t\big)\Big\rangle_{\varphi_2} \,
	g\big(|\vec{r}\,'|,\sphericalangle(\vec{r}\,');t\big) \label{eq:kirkwood-closure-force1}, \\
%%%
& \int d\varphi_2 \, g_1\Big( \vec{r}\,' \big| -\vec{r}, 0, 0, \varphi_2; t \Big) \notag \\
&\quad\quad = \Big\langle g\big(|\vec{r}\,'|,\varphi_2;t\big)\Big\rangle_{\varphi_2} \,
	g\big(|\vec{r}\,'+\vec{r}\,|,\sphericalangle(\vec{r}\,'+\vec{r});t\big) \label{eq:kirkwood-closure-force2} , 
\end{align}
where $\langle g(r,\varphi_2;t)\rangle_{\varphi_2}=(2\pi)^{-1}\int_{0}^{2\pi} d\varphi_2 g(r,\varphi_2;t)$ 
is an average over angles $\varphi_2$ holding the separation fixed and $\sphericalangle(\vec{r})$ denotes 
the angle enclosed by $\hat{e}_x$ and $\vec{r}$.

\subsection{Special case of hard disks}

\label{sec:special-case-hard-disks}

An important pair interaction is that of hard disks with only steric contributions. 
The reason is that short-range repulsive potentials can be mapped onto effective hard potentials 
with an effective particle diameter \cite{andersen_pra4_1971}. Thus, fundamental properties of 
systems dominantly governed by volume exclusion and packing 
can be studied and described by one unique model system of hard-core particles. 

In the special case of hard disks with diameter $\sigma$ ($1$ in our dimensionless units), 
the pair-interaction potential reads 
\begin{align}
u(r) &= \left\{ \begin{array}{lcl} \infty & \quad & r < 1 , \\ 0 & \quad & r > 1 . \end{array} \right. 
	\label{eq:hard-disk-potential}
\end{align}
In this case, the derivative of the pair potential simply becomes $u'(r)=-\delta(r-1)$, where $\delta$ 
denotes the Dirac-$\delta$ distribution. 
Accordingly, the force terms from Eqs.~(\ref{eq:force1b}) and (\ref{eq:force2b}) together 
with the Kirkwood closure from Eqs.~(\ref{eq:kirkwood-closure-force1}) and (\ref{eq:kirkwood-closure-force2}) 
lead to 
\begin{align}
\vec{F}_1(r,\theta;t)  
=& -\bar{\rho}
\int_{0}^{2\pi} d\theta' \hat{e}(\theta')
g\big(1,\theta';t\big) \notag \\
&\times 
\Big\langle g\Big(\big|
\hat{e}(\theta')
-r\hat{e}(\theta)
\big|,\varphi_2;t\Big)\Big\rangle_{\varphi_2} 
\label{eq:force1-C2} , \\
%%%%
	\vec{F}_2(r,\theta;t) =& 
	-\bar{\rho}
	\int_0^{2\pi} d\theta' \hat{e}(\theta')
\Big\langle g\big(1,\varphi_2;t\big)\Big\rangle_{\varphi_2} \notag \\
&\times	g\Big(
\big|
\hat{e}(\theta')
+r\hat{e}(\theta)
\big|, 
\sphericalangle\big(
\hat{e}(\theta')
+r\hat{e}(\theta)
\big);t\Big) \label{eq:force2-C2} , 
\end{align}
where $\hat{e}(\theta)=(\cos\theta,\sin\theta)$ denotes a unit vector in the direction of $\theta$. 
We further rewrite Eq.~(\ref{eq:eom5b}) by using the definition of the operators 
from Eqs.~(\ref{eq:gradient-operator}) and (\ref{eq:divergence-operator}) and by 
the pair potential from Eq.~(\ref{eq:hard-disk-potential}). 
Consequently, we find 
\begin{align}
	&	\partial_t g(r,\theta;t) \notag \\
=&\quad \frac{1}{r} \frac{\partial}{\partial r} \bigg( 
	r \hat{e}_r\cdot\vec{F}_1(r,\theta;t) 
	- r \hat{e}_r\cdot\vec{F}_2(r,\theta;t) \notag \\
&\quad\quad\quad\quad + r \, v_0 \hat{e}_r\cdot\hat{e}_1 
	+ 2 r \frac{\partial }{\partial r} \bigg) g(r,\theta;t) \notag \\
&+ 
	\frac{1}{r} \frac{\partial}{\partial \theta} \bigg( \hat{e}_\theta\cdot\vec{F}_1(r,\theta;t) 
	- \hat{e}_\theta\cdot\vec{F}_2(r,\theta;t) \notag \\
	&\quad\quad\quad\quad	+ v_0 \hat{e}_\theta\cdot\hat{e}_1 
	+ \frac{2}{r} \frac{\partial }{\partial \theta} \bigg) g(r,\theta;t) \notag \\
&+ 
	D_{\rm r}\frac{\partial^2 g(r,\theta;t)}{\partial \theta^2} 
	\label{eq:eom5c1}
\end{align}
for $r>1$. 
Since hard disks are not allowed to overlap, the flux in the radial direction 
at particle-particle contact must vanish with the no-flux condition 
\begin{align}
	& \Bigg( 
	\hat{e}_r\cdot\vec{F}_1(r,\theta;t) - \hat{e}_r\cdot\vec{F}_2(r,\theta;t) \notag \\
&\quad + \hat{e}_r\cdot\hat{e}_1 v_0 \Bigg) 
\times g(r,\theta;t)\bigg|_{r=1} 
= -\left.2\frac{\partial g(r,\theta;t)}{\partial r}\right|_{r=1} . 
\label{eq:no-flux}
\end{align}

\subsection{Simplified closure for hard disks}

\label{sec:simplification-hard-disks}

%+++++++++++++++++++++++++++++++++++++++
\begin{figure}
  \includegraphics[width=8.0cm]{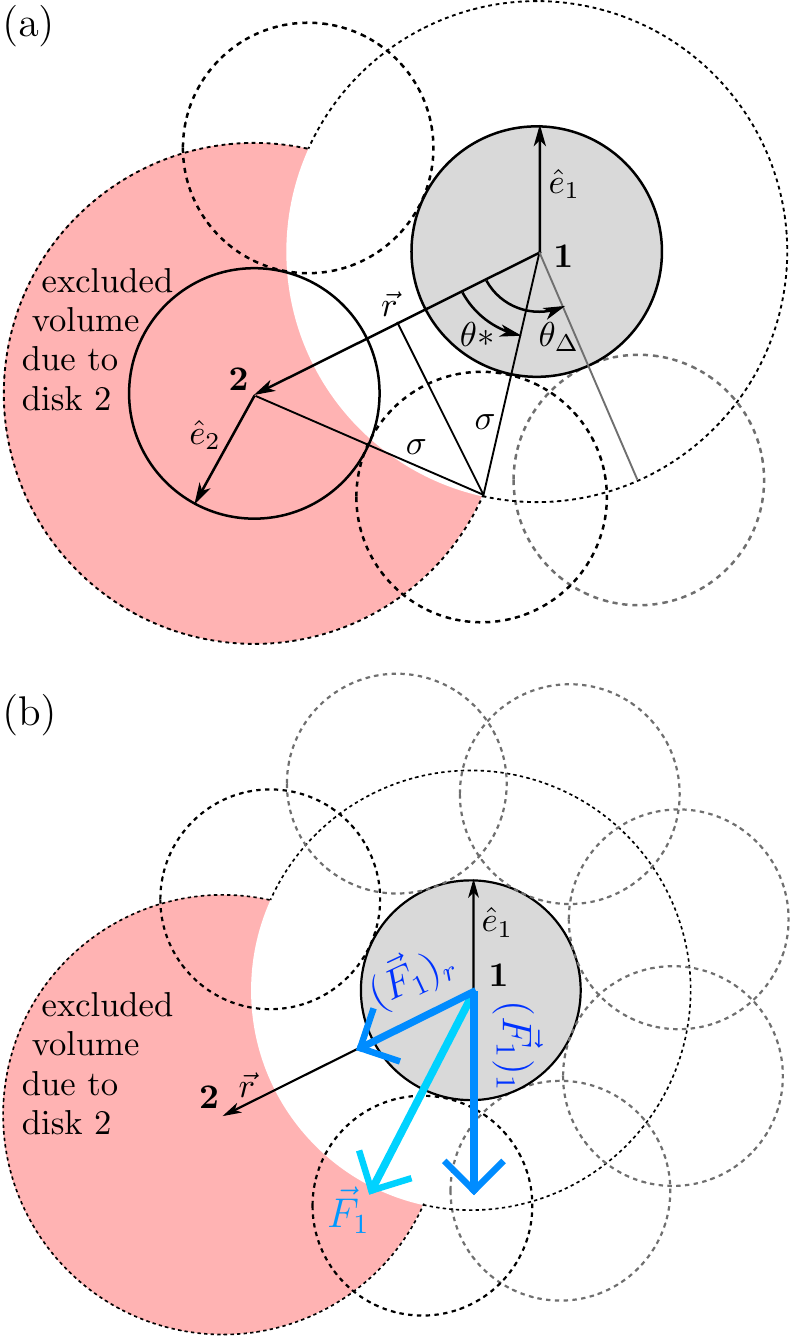} % 1000 x 821
	\caption{\label{fig:sketch-2particles}
	Sketch for two fixed hard disks labeled $1$ and $2$ and additional hard disks (dashed) 
	in contact with the first one. 
	The shaded area around particle 2 (red) is not accessible to a third particle due to the presence of the 
	second particle. 
	(a) Angles $\theta^*$ and $\theta_\Delta$ and (b) example of 
	the decomposition of the conditional force $\vec{F}_1$ acting on the first particle. The projected 
	components are $(\vec{F}_1)_r=(\hat{e}_r\cdot\vec{F}_1)\hat{e}_r$ and 
	$(\vec{F}_1)_1=(\hat{e}_1\cdot\vec{F}_1)\hat{e}_1$. }
\end{figure}
%+++++++++++++++++++++++++++++++++++++++

In order to achieve an analytical result, we will further simplify the conditional forces that we 
derived in the preceding section. It is known that fixing a single particle in bulk leads to a 
structured radial pair-distribution function. Now fixing a second particle 
at relative position $\vec{r}$ to 
the first particle (see Fig.~\ref{fig:sketch-2particles}) has a twofold outcome: 
On the one hand, it leads to a direct distribution around the two particles while, 
on the other hand, an indirect structure develops on top of the direct distribution due to 
the mutual influence of both fixed particles. For instance, these structures are discussed 
in a work on three-body correlations in passive systems \cite{zahn_prl91_2003}. 

These two contributions can also been understood from analyzing the force terms in 
Eqs.~(\ref{eq:force1-C2}) and (\ref{eq:force2-C2}), where two pair-distribution functions 
cause them, respectively. One contribution stems from the interplay between the third and the 
fixed particle, on which the respective force $\vec{F}_i$ is acting, while 
the second contribution arises from the interplay between the third and the remaining 
second particle. This situation is sketched in Fig.~\ref{fig:sketch-2particles}(a) for the fixed 
particle having the index $i=1$. 

To give an example, we first discuss a similar situation where two hard disks are in contact with 
a third one. This system has been studied by Attard, who proposed an adjusted Kirkwood approximation 
as a reasonably good closure \cite{attard_mp74_1991}. The system he studied corresponds to 
the situation shown in Fig.~\ref{fig:sketch-2particles} for $|\vec{r}\,|=1$, when the second and third 
particles both are in contact with the particle labeled by $1$. The third particle can move along the surface 
of particle $1$ and its position can be parametrized by the enclosed angle $\theta_{\Delta}$. 
The closure Attard proposed reads \cite{attard_mp74_1991} 
\begin{align}
g_1\big(1,1,\cos\theta_\Delta\big) 
	&= g(1) g(1) \Bigg(1+\frac{g\big(s(\theta_\Delta)\big)-1}{2} \Bigg) , 
	\label{eq:adjusted-kirkwood-closure} \\
	s(\theta_\Delta) &= \left\{ \begin{array}{lcl}
	1 + 1(\theta_\Delta-\theta^*) & \quad & \theta_\Delta \leq \pi , \\
	1 + 1(2\pi-\theta_\Delta-\theta^*) & \quad & \theta_\Delta > \pi , 
	\end{array} \right.
\end{align}
which is valid for $\theta^* \leq \theta_\Delta \leq 2\pi-\theta^*$ with $\theta^*=\arccos(1/2)=\pi/3$. 
For other values of $\theta_\Delta$ the probability of finding a particle vanishes and 
$g_1(1,1,\cos\theta_\Delta)=0$, because particles 2 and 3 are not allowed to overlap. 
In this approximation, the separation between the two particles 2 and 3 
is not measured along a straight line but along the surface of the first particle. 
The angle $\theta^*=\pi/3$ denotes the limiting case when both particles 2 and 3 are in contact. 
Note that for separations $r=|\vec{r}\,|>1$ this angle gets smaller dependent on the separation $r$. 

In our theory, the approximation proposed by Attard \cite{attard_mp74_1991} relates 
to the pair-distribution 
function between particles 2 and 3. This function can be split into two contributions: 
One simply originates from the excluded volume that the third particle cannot access due to the 
presence of the second particle; the other contribution stems from the indirect part of the pair 
distribution between particles 2 and 3. 
Considering the closure by Attard \cite{attard_mp74_1991} in Eq.~(\ref{eq:adjusted-kirkwood-closure}), 
neglecting the second contribution would correspond to approximating the term $(1+\dots)$ in 
Eq.~(\ref{eq:adjusted-kirkwood-closure}) as $1$. 
In our theory, we would have to replace the respective pair-distribution function $g$ in the 
conditional forces in Eqs.~(\ref{eq:force1-C2}) and (\ref{eq:force2-C2}) by a 
spherical step function 
\begin{align}
g(r,\theta) \quad \to \quad 
\left\{ \begin{array}{lcl} 0 & \quad & r < 1 , \\ 1 & \quad & r \geq 1 . \end{array} \right. 
\label{eq:assumption-of-spherical-step-function}
\end{align}
When we apply this {\it{simplification}} 
to the respective second function $g$ in Eqs.~(\ref{eq:force1-C2}) and (\ref{eq:force2-C2}), 
the conditional forces simplify to 
\begin{align}
	\vec{F}_1(r,\theta;t) =& 
	-\bar{\rho} \int_{0}^{2\pi} d\varphi 
	\begin{pmatrix}\cos(\varphi)\\ \sin(\varphi)\end{pmatrix} g(1,\varphi;t) \notag \\
	&
	+\bar{\rho} \int_{\theta-\theta^*}^{\theta+\theta^*} d\varphi 
	\begin{pmatrix}\cos(\varphi)\\ \sin(\varphi)\end{pmatrix} g(1,\varphi;t) , 
	\label{eq:force1fb} \\
%%%	
	\vec{F}_2(r,\theta;t) =& 
	-\bar{\rho} \int_{0}^{2\pi} d\varphi 
	\begin{pmatrix}\cos(\varphi)\\ \sin(\varphi)\end{pmatrix} \big\langle g(1,\varphi_2;t)\big\rangle_{\varphi_2} 
	\notag \\
	&
	+\bar{\rho} \int_{\theta+\pi-\theta^*}^{\theta+\pi+\theta^*} d\varphi 
	\begin{pmatrix}\cos(\varphi)\\ \sin(\varphi)\end{pmatrix} \big\langle g(1,\varphi_2;t)\big\rangle_{\varphi_2} . 
	\label{eq:force2fb}
\end{align}
The limiting $r$-dependent angle $\theta^*$ that spans the excluded area [see 
Fig.~\ref{fig:sketch-2particles}(a)] reads 
\begin{align}
\theta^* = \theta^*(r) = 
\left\{ \begin{array}{lcl} \arccos(r/2) & \quad & 1 \leq r \leq 2 , \\
0 & \quad & r > 2 . \end{array} \right. \label{eq:excluded-volume-angle} 
\end{align}

Now we can identify two main contributions to the conditional forces $\vec{F}_i$ in our theory. 
We can understand their origin from Fig.~\ref{fig:sketch-2particles}(b), where an exemplary  
force $\vec{F}_1$ is constructed from two components. 
The {\it{first component}} 
$(\vec{F}_1)_r=(\hat{e}_r\cdot\vec{F}_1)\hat{e}_r$ acts along the separation vector 
$\vec{r}$. It originates from the excluded volume due to disk 2 such that the surrounding third 
particles on average push the first particle (approximately) in the direction of the excluded volume. 
This component is expected to vanish for large separations $r=|\vec{r}\,|$. 
In our theory, we can see this behavior from Eqs.~(\ref{eq:force1fb}) and (\ref{eq:force2fb}). 
If $g(r,\theta;t)$ were homogeneous in the angle $\theta$, the respective second terms on 
the right-hand sides of Eqs.~(\ref{eq:force1fb}) and (\ref{eq:force2fb}) would point exactly along 
the direction of the separation vector $\vec{r}$. 
The {\it{second component}} 
$(\vec{F}_1)_1=(\hat{e}_1\cdot\vec{F}_1)\hat{e}_1$ along the direction 
$\hat{e}_1$ of self-propulsion of the first particle clearly originates from collisions with surrounding 
third particles. This component is expected to be independent of $r$ at 
large separations and to vanish only in the limit of vanishing propulsion speed $v_0$. 
Moreover, the function $g(r,\theta;t)$ is symmetric in the angle $\theta$, i.e., $g(r,\theta;t)=g(r,-\theta;t)$. 
For this reason, the first term on the right-hand side of Eq.~(\ref{eq:force1fb}) points exactly 
along the orientation $\hat{e}_1$ of the first particle, while the first term on the right-hand side of 
Eq.~(\ref{eq:force2fb}) vanishes. 
In conclusion, the two main directions of the contributions to the conditional forces $\vec{F}_i$, 
as shown in Fig.~\ref{fig:sketch-2particles}(b), are the direction of the (normalized) separation vector 
$\hat{e}_{\rm r}=\vec{r}/|\vec{r}\,|$ between both tagged particles and the direction of self-propulsion 
$\hat{e}_1$ of the first particle.

\subsection{Expansion of the pair-distribution function}
\label{sec:expansion-g}

In this section we derive analytic expressions for the conditional forces, for the effective swimming 
speed, and for some properties of the pair-distribution function in systems of ABPs. We further define 
parameters to characterize systems of ABPs following previous work. 
According to the identification of the two main directions in the preceding section, 
we will study the projections of the conditional forces onto those directions 
and derive explicit terms from our theory. 
In this context, we are solely interested in steady-state solutions of Eq.~(\ref{eq:eom5c1}) and 
for this reason we will skip the parameter $t$ throughout the remaining part of our work. 

To achieve analytical expressions for the conditional forces $\vec{F}_i$, we expand the 
pair-distribution function $g(r,\theta)$ in Fourier modes by 
\begin{align}
	g(r,\theta) &= \sum_{k=0}^{\infty} g_k(r) \cos(k \theta) . \label{eq:g-expansion}
\end{align}
We discuss details on the full expansion in the Appendix. 
When we neglect higher Fourier modes with $k>1$, we find the resulting projections of the conditional force 
$\vec{F}_1$ onto $\hat{e}_r$ and $\hat{e}_\theta$ with 
\begin{align}
	\hat{e}_r\cdot\vec{F}_1(r,\theta) =& 
	2 \bar{\rho} g_0(1) \sin(\theta^*) \notag \\
	& -\bar{\rho} g_1(1) \bigg(\pi-\theta^*-\sin(\theta^*)\frac{r}{2}\bigg) \cos(\theta) , 
	\label{eq:proj-f1-er} \\
%%%%%%%%%%%%%%%
	\hat{e}_\theta\cdot\vec{F}_1(r,\theta) =& 
	\bar{\rho} g_1(1) \bigg(\pi-\theta^*+\sin(\theta^*)\frac{r}{2}\bigg) \sin(\theta) . 
\end{align}
The limiting angle $\theta^*(r)$ that spans the excluded area due to the presence of particle 2 
has been defined in Eq.~(\ref{eq:excluded-volume-angle}). The angle is shown in 
Fig.~\ref{fig:sketch-2particles} and, for completeness, we give $\sin(\theta^*)=\sqrt{1-(r/2)^2}$. 
The orientation of the first particle 
is given by $\hat{e}_1 = \cos(\theta)\hat{e}_r - \sin(\theta)\hat{e}_\theta$ 
such that we also find 
\begin{align}
	\frac{\hat{e}_1\cdot\vec{F}_1(r,\theta)}{\bar{\rho}} &= 
		f_{\rm a}(r) + f_{\rm b}(r) \cos(\theta) + f_{\rm c}(r) \cos(2\theta) , \label{eq:proj-f1-e1} \\
	f_{\rm a}(r) &= g_1(1) \big(\theta^*(r)-\pi\big) , \label{eq:coef-fa} \\
	f_{\rm b}(r) &= 2 g_0(1) \sin\big(\theta^*(r)\big) , \label{eq:coef-fb} \\
	f_{\rm c}(r) &= g_1(1) \sin\big(\theta^*(r)\big)\frac{r}{2} . \label{eq:coef-fc}
\end{align}
For large separations $r\geq 2$ the function $\theta^*$ vanishes and we find 
$\hat{e}_1\cdot\vec{F}_1(r,\theta) = -\bar{\rho} \pi g_1(1)$. This finding agrees with our expectation 
from the preceding section, where we discussed that, in this limit, the second particle does not 
affect the contribution of third particles on the first one anymore. Consequently, we find a constant 
force along the direction of propulsion of the first particle. 

In previous work, Speck {\it at al.} analyzed the anisotropy of the pair-distribution function for 
active colloidal disks by studying an anisotropy parameter of this function 
\cite{bialke_epl103_2013,bialke_jncs407_2015,speck_jcp142_2015}. 
Following the definition of this parameter $\zeta_1$ in previous work \cite{bialke_epl103_2013}, 
we define the first two moments 
\begin{align}
	\zeta_0 &= - \int_0^\infty dr \, r \, u'(r) \int_0^{2\pi} d\theta g(r,\theta) , \label{eq:zeta0} \\
	\zeta_1 &= - \int_0^\infty dr \, r \, u'(r) \int_0^{2\pi} d\theta \cos(\theta) g(r,\theta) . \label{eq:zeta1}
\end{align}
These parameters can easily be extracted from simulations and could have the role of order parameters 
in the description of systems of ABPs and their states. 
For hard disks, we find relations between these parameters and the prefactors $g_i(1)$ 
in the expansion from Eq.~(\ref{eq:g-expansion}) at particle contact ($r=1$) that read 
\begin{align}
\zeta_0 &= 2\pi g_0(1) , \label{eq:rel-zeta-g0} \\
\zeta_i &= \pi g_i(1)  \quad \forall i\geq 1 . \label{eq:rel-zeta-gi}
\end{align}
For almost hard potentials, we will discuss deviations from this equalities in the following sections. 

Further insight into our theory is gained by considering the flux $\vec{j}$ that follows from 
$\partial_t g(r,\theta;t)=-\vec{\nabla}\cdot\vec{j}$ both at 
particle contact ($r=1$) and for infinite particle separation ($r\to\infty$). 
In the case of particle contact, we can combine the expansion (\ref{eq:g-expansion}) 
and the no-flux condition (\ref{eq:no-flux}), as shown in the Appendix. 
In the limit of vanishing propulsion speed $v_0\to 0$, where all $g_k$ for $k>0$ vanish, 
we find $g_0'(1)=-\sqrt{3}g_0(1)g_0(1)\bar{\rho}$ [see Eq.~(\ref{eq:no-flux-res1})] 
as an analytical result for passive systems. 
In the case of large particle separations $r\to\infty$, both tagged particles are uncorrelated and 
the flux in the moving reference system 
is simply given by the effective swimming speed $v$ of the tagged first particle in the opposite direction 
of propulsion, i.e., $\vec{j}=-v\hat{e}_1$. In this limit, our theory in Eq.~(\ref{eq:proj-f1-e1}) 
predicts a flux $\bar{\rho}\zeta_1\hat{e}_1-v_0\hat{e}_1$ such that we find the relation 
\begin{align}
v &= v_0 - \bar{\rho}\zeta_1 
\end{align}
in accord with previous work \cite{bialke_epl103_2013}.

%%%%%%%%%%%%%%%%%%%%%%%%%%%%%%%%%%%%%%%%%%%%%%%%%%%%%%%%%%%%%%%%%%%%%%%%%%%%%%%%%%%%%%%%%%%%%%%%%%
%%%%%%%%%%%%%%%%%%%%%%%%%%%%%%%%%%%%%%%%%%%%%%%%%%%%%%%%%%%%%%%%%%%%%%%%%%%%%%%%%%%%%%%%%%%%%%%%%%

\section{Results}

The first main result of this work is the analytic theory for the microscopic structure around a 
tagged particle in suspensions of active Brownian particles that we derived in the preceding 
section. For instance, the 
theory describes the conditional forces $\vec{F}_k$ as defined in Eq.~(\ref{eq:force-3particle}). 
In order to achieve a more detailed picture and to apply and test our theory, we also perform 
Brownian dynamics (BD) simulations that we describe in this section first. 
Then we draw a direct comparison between our theoretical predictions and our results from 
simulations. In a third step, we test our theory for a general pair-distribution function, i.e., 
without skipping higher modes in the expansion from Sec.~\ref{sec:expansion-g}. For this purpose, 
we solve Eq.~(\ref{eq:eom5c1}) numerically and compare its solutions to the results from our simulations.

\subsection{Brownian dynamics simulations}
\label{sec:bd-sim}

%+++++++++++++++++++++++++++++++++++++++
\begin{figure*}
  \includegraphics[width=16.0cm]{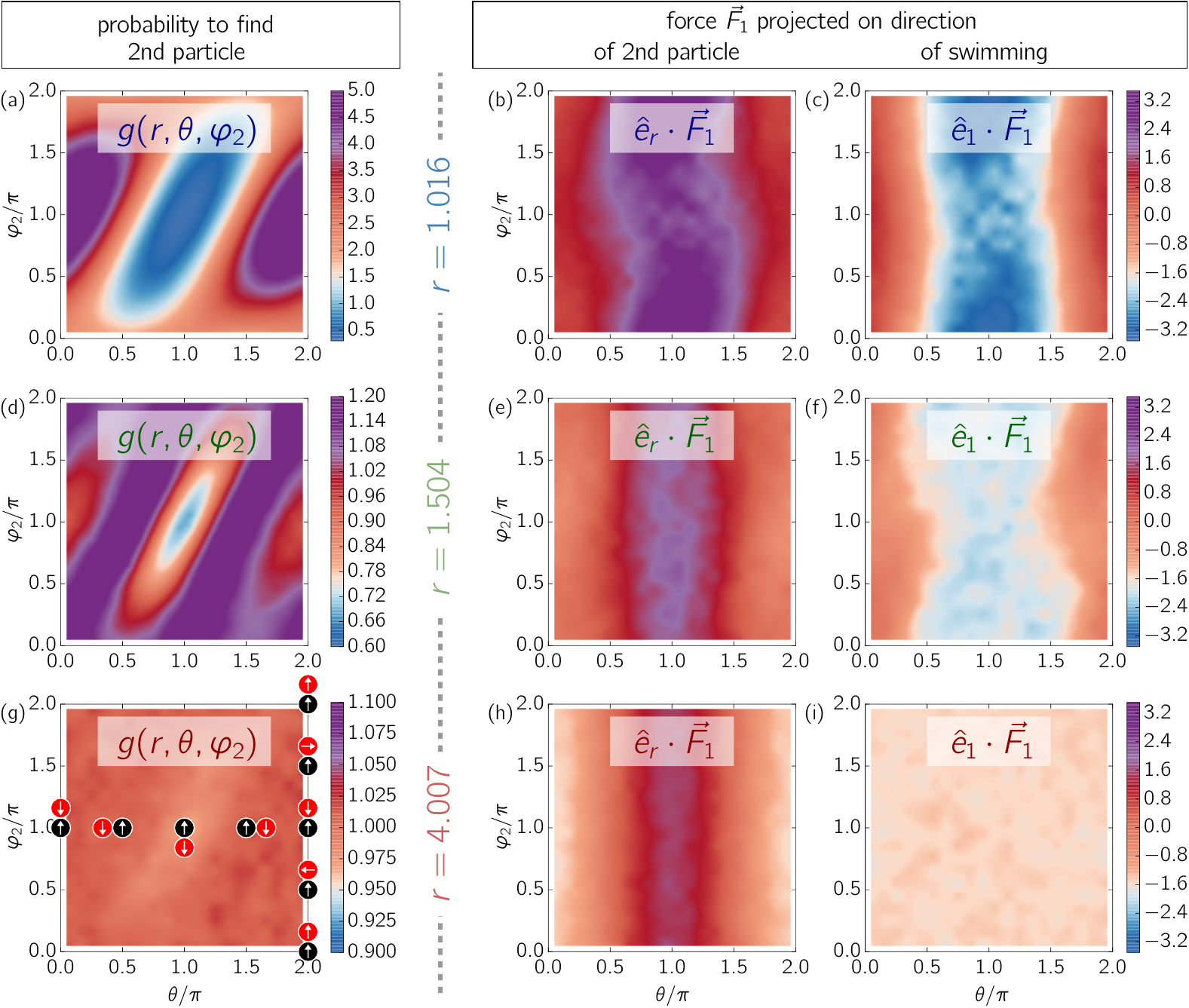}
	\caption{\label{fig:force2d}
	Pair distribution $g(r,\theta,\varphi_2)$ (first column) and conditional force $\vec{F}_1(r,\theta,\varphi_2)$ 
	(second and third columns) from BD simulations at number density $\bar{\rho}=0.3$ 
	and propulsion speed $v_0/d_{\rm eff}=5$. 
	The first column shows the distribution of a second particle around the first one, 
	the second column shows the projection of $\vec{F}_1$ onto the radial direction $\hat{e}_{\rm r}$, 
	and the third column shows the projection of $\vec{F}_1$ onto the orientation $\hat{e}_1$. 
	The relative position of the second particle with respect to the first one is 
	given by the separations (a)-(c) $r\approx 1$, (d)-(f) $r\approx 1.5$, and (g)-(i) $r\approx 4$ 
	and by the angle $\theta$, where $\theta=0$ corresponds to the position in front of the tagged first 
	particle as sketched in Fig.~\ref{fig:sketch-coordinates}(b). The relative orientation of the second 
	particle with respect to the first one is given by $\varphi_2$. To help interpret these plots both the 
	relative position $\theta$ and the orientation $\varphi_2$ are sketched in (g) for certain 
	settings of particles 1 (black) and 2 (red) at the corresponding position in the plot. 
	}
\end{figure*}
%+++++++++++++++++++++++++++++++++++++++

%+++++++++++++++++++++++++++++++++++++++
\begin{figure*}
  \includegraphics[width=14.0cm]{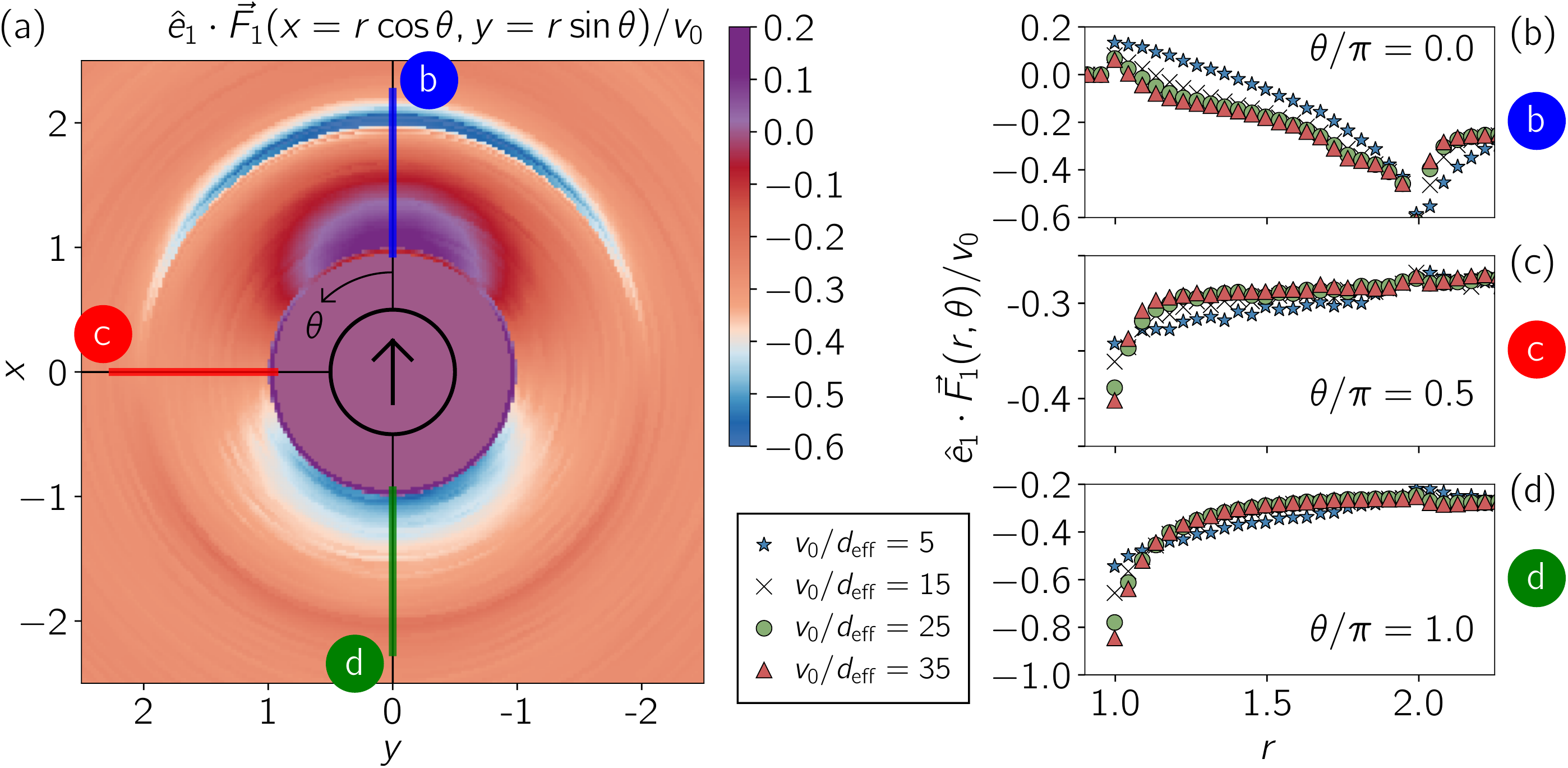}
	\caption{\label{fig:force1_e1}
		Projected conditional force $\hat{e}_1\cdot\vec{F}_1$ on a tagged particle dependent on the relative 
		position $(r,\theta)$ of a second particle as obtained from our BD simulations with a 
		number density $\bar{\rho}=0.3$. 
		Data are shown for (a) propulsion speed $v_0/d_{\rm eff}=5$ and 
		(b)-(d) speeds $5$, $15$, $25$, and $35$. As indicated in (a), the data of (b)-(d) are shown along the 
		cutting lines (b) along the positive $x$ axis (in the direction of propulsion), 
		(c) along the positive $y$ axis (equal to the negative $y$ axis), 
		and (d) along the negative $x$ axis. }
\end{figure*}
%+++++++++++++++++++++++++++++++++++++++

We simulate $N=4096$ two-dimensional Brownian 
swimmers interacting via the repulsive short-range Weeks-Chandler-Andersen (WCA) potential 
\begin{align}
	u_{\rm WCA}(r)
	&= 4\epsilon \bigg[ \Big(\frac{\lambda}{r}\Big)^{12}-\Big(\frac{\lambda}{r}\Big)^{6} + \frac{1}{4} \bigg]
	\label{eq:wca-potential}
\end{align}
for $r\leq r_c=2^{1/6}\lambda$ and zero otherwise. 
We employ overdamped dynamics as described in Eq.~(\ref{eq:eom}), 
where $\Delta t=t-t'$ is the time step. The orientation $\varphi$ 
undergoes free rotational diffusion with a diffusion constant $D_r=3D_0/\delta^2$, where $\delta$ is 
the particle diameter. We set $\delta$ equal to the effective diameter 
$\delta=d_{\rm eff}\lambda$, computed by the 
Barker-Henderson approximation \cite{barker_jcp47_1967,andersen_pra4_1971}. 
The energy is scaled by a bath temperature $k_{\rm B} T$. 
The repulsive strength $\epsilon$ of the potential is set to $100 k_{\rm B}T$, 
which results in $d_{\rm eff}=1.10688$. The time step is set to $2\times 10^{-6}\lambda^2/D_0$. 

To obtain the conditional force $\vec{F}_1$ from our BD simulations, we have chosen an equidistant 
binning of $2\pi/20$ for each angle $\theta$ and $\varphi_2$, respectively, and $5/500$ for the separation 
$r$. To check consistency, we additionally have calculated distribution functions at a higher resolution 
$2\pi/80$ and $2/1000$; the calculation of distribution functions is less time consuming than the calculation 
of the three-body forces. We found almost no deviations between the data for both resolutions. 

Figure \ref{fig:force2d} shows data obtained for a number density $\bar{\rho}=0.3$ and 
a propulsion speed $v_0/d_{\rm eff}=5$. 
%  5 ->  5.5344
% 10 -> 11.0688
% 15 -> 16.6032
% 20 -> 22.1376
% 25 -> 27.672
% 30 -> 33.2064
% 35 -> 38.7408
For each of the 80000 snapshots that we analyzed after the system was equilibrated, we successively tagged two 
particles and summed up the force contributions of all remaining particles onto the first one. 
Figures \ref{fig:force2d}(a), \ref{fig:force2d}(d), and \ref{fig:force2d}(g) show the distribution 
of second particles around the tagged first particle as used for our analysis. 
The axes correspond to the angular position $\theta$ of the second particle in relation to the propulsion 
direction and position of the tagged first particle, as well as to the orientation 
angle $\varphi_2$ of the second particle relative to 
that of the first one. This situation is also illustrated in Fig.~\ref{fig:sketch-coordinates}(b). 
In addition, we sketch a visualization of the different relative positions and orientations in 
Fig.~\ref{fig:force2d}(g) for selected settings. Here the position of the second (red) particle relative 
to the tagged (black) particle changes along the $\theta$ axis and the relative orientation of the second 
particle (direction of arrows) changes along the $\varphi_2$ axis. 
In the different rows of Fig.~\ref{fig:force2d}, 
we show data for three absolute separations $r\approx 1$ [Figs.~\ref{fig:force2d}(a)-\ref{fig:force2d}(c)], 
$r\approx 1.5$ [Figs.~\ref{fig:force2d}(d)-\ref{fig:force2d}(f)], 
and $r\approx 4$ [Figs.~\ref{fig:force2d}(g)-\ref{fig:force2d}(i)]. 
For small separations of particles 1 and 2 we observe a much higher probability 
of finding a second particle in front [spot at $(\theta/\pi,\varphi_2/\pi)=(0,1)$ in 
Fig.~\ref{fig:force2d}(a)] than 
behind [spot at $(1,1)$] the first particle when both particles have opposite orientations. 
When both particles have the same orientation (when they move together), second particles seem to be 
distributed uniformly around the first particle [$\varphi_2=0$ in Fig.~\ref{fig:force2d}(a)]. 
For increasing separation $r$ between both particles, the observed spots in the distribution get 
less pronounced [see Fig.~\ref{fig:force2d}(d)] 
and vanish completely in the uniform distribution in Fig.~\ref{fig:force2d}(g). 

Figures \ref{fig:force2d}(b), \ref{fig:force2d}(e), and \ref{fig:force2d}(h) and 
Figs~\ref{fig:force2d}(c), \ref{fig:force2d}(f), and \ref{fig:force2d}(i) show the projection 
of the conditional force $\vec{F}_1$ on the direction of the separation $\vec{r}$ between both 
tagged particles and on the direction of propulsion of the first particle, respectively. 
The choice of these directions is motivated by the main directions that can be identified in 
the conditional forces in Eqs.~(\ref{eq:force1fb}) and (\ref{eq:force2fb}) and we have discussed 
their origin and expected values using Fig.~\ref{fig:sketch-2particles}(b) in Sec.~\ref{sec:expansion-g}. 
In accordance with these expectations, the value of the projection $\hat{e}_1\cdot\vec{F}_1$ 
becomes constant for large separations as shown in Fig.~\ref{fig:force2d}(i), because $\vec{F}_1$ 
becomes parallel to $\hat{e}_1$. Our theory in Sec.~\ref{sec:expansion-g} 
even predicts the value $\hat{e}_1\cdot\vec{F}_1=-\bar{\rho}\zeta_1$, 
which perfectly fits to a $\zeta_1\approx 5.0$ that corresponds to the system analyzed in Fig.~\ref{fig:force2d}. 
The uniform distribution further is confirmed by the projection $\hat{e}_r\cdot\vec{F}_1$ in 
Fig.~\ref{fig:force2d}(h), which shows a cosinelike 
dependence on $\theta$ as expected from the definition of $\hat{e}_1=(1,0)$ and 
$\hat{e}_r=(\cos\theta,\sin\theta)$ in Sec.~\ref{sec:symmetries-parametrization}. At 
small separations $r$ the excluded-volume effect of the second particle becomes important too. 
For instance, Fig.~\ref{fig:force2d}(c) shows that at $\theta=\pi$ 
the constant value of approximately $-1.6$ from Fig.~\ref{fig:force2d}(i) has doubled to a value of 
around $-3.2$. 
In this situation, the second particle is located behind the first one such that any third particle 
likely pushes the first one from ahead due to the excluded volume. This component adds to the 
collision effect due to the propulsion of the particle. In comparison, at $\theta=0$ the second 
particle is located in front of the first one such that the force due to the excluded volume 
pushes the particle from behind. However, the excluded volume of the second particle in front of the 
first particle at the same time prevents collisions with third particles such that the absolute value 
of the projected force in Fig.~\ref{fig:force2d}(c) at $\theta=0$ is only half of the value at $\theta=\pi$.  

Moreover, the results in Fig.~\ref{fig:force2d} illustrate that the dependence 
of the force $\vec{F}_1$ on the orientation $\varphi_2$ of the second particle is weak in 
comparison to the relative position of the second particle. In particular, 
Figs.~\ref{fig:force2d}(c), \ref{fig:force2d}(e), \ref{fig:force2d}(h), and \ref{fig:force2d}(i) 
show almost no dependence on the orientation $\varphi_2$, while Figs.~\ref{fig:force2d}(b) and 
\ref{fig:force2d}(f) show only minor dependences. 
Interestingly, when particles 1 and 2 are in contact, the dependence on the orientation $\hat{e}_2$ 
of the second particle is stronger for the projection $\hat{e}_r\cdot\vec{F}_1$ 
[Fig.~\ref{fig:force2d}(b) vs. Fig.~\ref{fig:force2d}(c)], 
while at intermediate separations it is stronger for the projection $\hat{e}_1\cdot\vec{F}_1$ 
[Fig.~\ref{fig:force2d}(f) vs. Fig.~\ref{fig:force2d}(e)]. 
Furthermore, we could connect the strength of inhomogeneities in 
the projection of the force onto $\hat{e}_r$ with the strength of its component due to collisions, 
if we study the limit of large separations in Figs.~\ref{fig:force2d}(h) and \ref{fig:force2d}(i). 
If this connection would 
also hold at small separations, the orientation of the second particle would be most important 
for the force component due to collisions at particle contact ($r\approx 1$) and for the component 
due to excluded volume at intermediate separations $r$. 

From Fig.~\ref{fig:force2d} we could conclude that the resulting force and its anisotropy are weakened 
when the separation $r$ between the two particles $1$ and $2$ is increased. For this reason, we 
study the dependence of the conditional force $\vec{F}_1$ on the separation $r$ between the two 
particles in more detail using Fig.~\ref{fig:force1_e1}. To obtain the data shown in 
Fig.~\ref{fig:force1_e1} we have averaged over the orientation $\varphi_2$ of the second particle, 
which we previously have seen to have only a minor impact on the force $\vec{F}_1$. 
The plot in Fig.~\ref{fig:force1_e1}(a) is supported by the three right plots in 
Figs.~\ref{fig:force1_e1}(b)-\ref{fig:force1_e1}(d) that show data along the marked cutting 
lines b, c, and d. 
These supporting plots also present data for additional propulsion speeds. 
The data clearly show an exception from a monotonic decay of the force strength with increasing 
separation $r$ at a separation of $r\approx 2$: For a second particle located ahead of the 
tagged particle, the conditional force shows a strong dip. This dip exists because 
a third particle exactly fits in between particles 1 and 2 when the second particle is located 
at $r\gtrsim 2$. This third particle would block the self-propelled particle 1 and create a strong 
force slowing down the movement of particle 1. 
A similar but less pronounced reaction would also be expected at around $r\approx 3$ in situations 
of four particles in a row. Indeed, we have found such settings in our simulations 
as shown in Fig.~\ref{fig:sketch-coordinates}(c). 
Note that the dip at $r\gtrsim 2$ and those expected at higher locations are not described by our 
simplified theory because we have neglected an additional structure between particles 2 and 3 
in our assumption from Eq.~(\ref{eq:assumption-of-spherical-step-function}). 

We mention that the force term $\vec{F}_1$ overall seems to depend on the propulsion speed linearly, 
as we see from the collapse of the curves in Figs.~\ref{fig:force1_e1}(b)-\ref{fig:force1_e1}(d). 
We observe the largest deviations from this linear dependence in the front of the tagged particle 
[at $\theta=0$, stars in Fig.~\ref{fig:force1_e1}(b)] and at small propulsion speeds.

\subsection{Test of the theoretical predictions}
\label{sec:test-theory}

%+++++++++++++++++++++++++++++++++++++++
\begin{figure}
  \includegraphics[width=8.5cm]{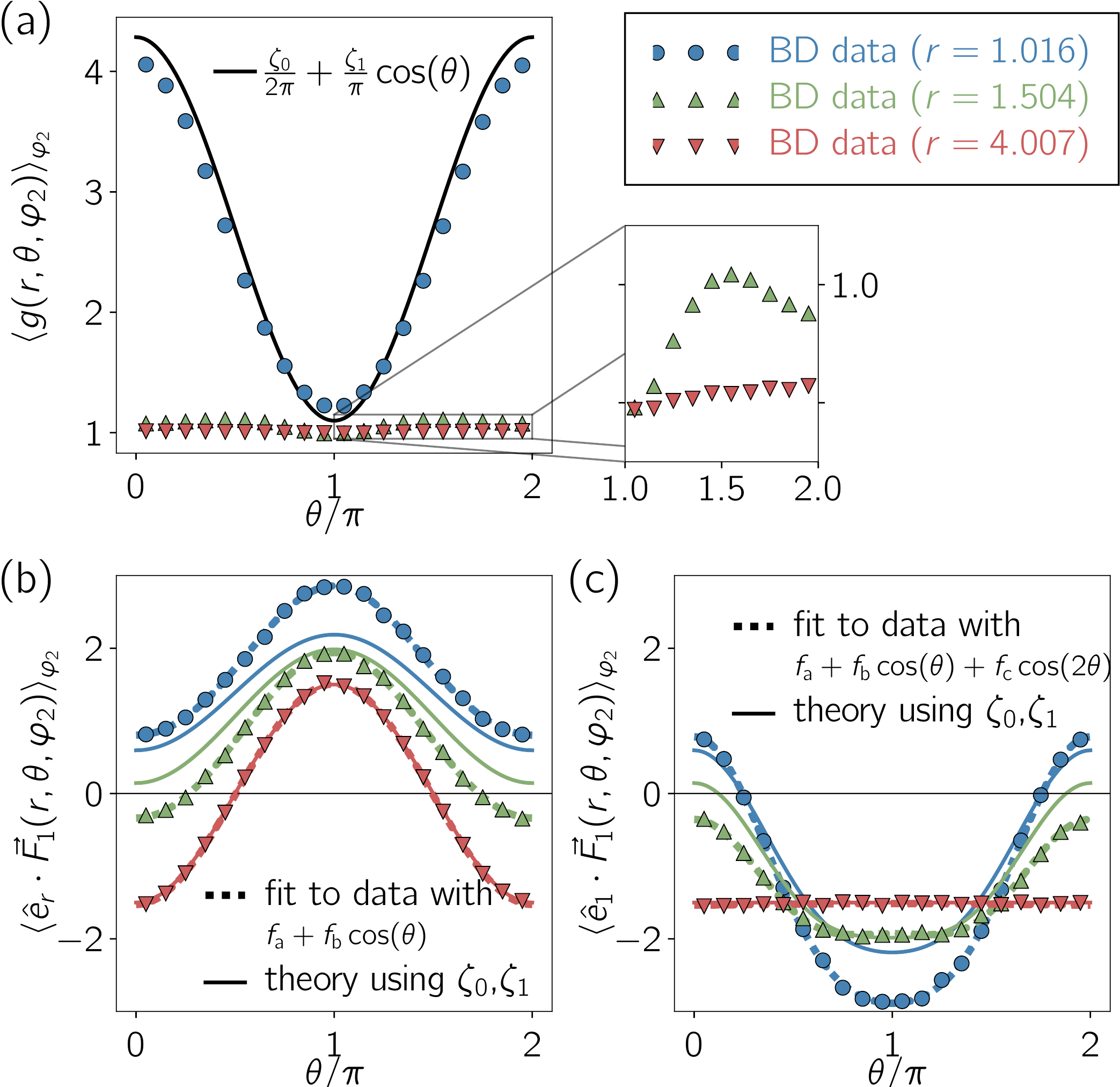}
	\caption{\label{fig:force2d-averaged} 
	(a) Pair distribution $\langle g(r,\theta,\varphi_2) \rangle_{\varphi_2}$ and (b) and (c) projected 
	conditional force $\vec{F}_1(r,\theta,\varphi_2)$ as shown in Fig.~\ref{fig:force2d} 
	($v_0/d_{\rm eff}=5$ and $\bar{\rho}=0.3$), but 
	averaged over the angle $\varphi_2$ of the relative orientation of the second particle. 
	The plots show the averaged simulation data from Fig.~\ref{fig:force2d} (symbols), 
	least-squares fits to the data in (b) and (c) as noted in the respective legend 
	(dotted lines), and theoretical predictions (solid lines) from (a) Eq.~(\ref{eq:g-expansion}), (b) 
	Eq.~(\ref{eq:proj-f1-er}), and (c) Eq.~(\ref{eq:proj-f1-e1}). 
	For the theoretical predictions we use the parameters $\zeta_0=2\pi g_0(1)$ and $\zeta_1=\pi g_1(1)$, 
	which we have calculated from our simulations via Eqs.~(\ref{eq:zeta0}) and 
	(\ref{eq:zeta1}). 
	}
\end{figure}
%+++++++++++++++++++++++++++++++++++++++

As a next step, we test our theoretical predictions from Sec.~\ref{sec:expansion-g} by comparing 
them to our BD simulations. In particular, we are interested in the collapse 
of data that we have observed in the preceding section in Fig.~\ref{fig:force1_e1}. 
We follow two routes for our comparison. First, we extract the parameters $\zeta_i$ 
from the pair-distribution functions in our simulations and discuss them in the context of our theory. 
Second, we compare the theoretical form of the projection $\hat{e}_1\cdot\vec{F}_1$ in 
Eq.~(\ref{eq:proj-f1-e1}) to the projected force measured in our simulations and shown in 
Fig.~\ref{fig:force2d}. Note that along the second route we also extract the parameters $g_i(1)$ 
that are contained in the prefactors $f_{\rm a}$, $f_{\rm b}$, and $f_{\rm c}$ of Eq.~(\ref{eq:proj-f1-e1}). 
For hard disks we found the relations from Eqs.~(\ref{eq:rel-zeta-g0}) and (\ref{eq:rel-zeta-gi}) 
between the $\zeta_i$ and the $g_i(1)$. 

First, we use Eqs.~(\ref{eq:zeta0}) and (\ref{eq:zeta1}) to extract the parameters $\zeta_0$ and $\zeta_1$ 
from our simulation results that we have shown in Fig.~\ref{fig:force2d}. We find 
$\zeta_0\approx16.9$ and $\zeta_1\approx5.0$. To allow a comparison to our theory, we average the data 
shown in Fig.~\ref{fig:force2d} over the orientation $\hat{e}_2$ of the second particle, because this 
parameter has been averaged out in our theory too. The averaged data are presented in 
Fig.~\ref{fig:force2d-averaged}. In Fig.~\ref{fig:force2d-averaged}(a) we show 
the resulting pair-distribution functions $g(r,\theta)$ from Figs.~\ref{fig:force2d}(a), 
\ref{fig:force2d}(d), and \ref{fig:force2d}(g) together with the predicted function 
\begin{align}
g(1,\theta) &= \frac{\zeta_0}{2\pi}+\frac{\zeta_1}{\pi}\cos(\theta)
\end{align}
at contact that follows from the extracted $\zeta_0$ and $\zeta_1$ via the first 
terms of the expansion in 
Eq.~(\ref{eq:g-expansion}). We observe minor deviations between the 
simulation data at $r=1.016$ and the theoretical prediction using $\zeta_0$ and $\zeta_1$, because 
the $\zeta_i$ correspond to the $g_i(1)$ of hard disks (see also Sec.~\ref{sec:pair-distribution-function}). 
Overall, the expansion of the pair-distribution function with only two modes captures the simulation data 
very well at $r\approx 1$ and at large $r$, but it cannot capture the additional modes that occur 
at intermediate separations $r\approx1.5$, which we can see in the inset of Fig.~\ref{fig:force2d-averaged}(a). 

In accord with this finding on the pair-distribution function, we also observe the strongest deviations 
at intermediate separations $r\approx1.5$ between the theoretical predictions and simulation data in 
Figs.~\ref{fig:force2d-averaged}(b) and \ref{fig:force2d-averaged}(c), 
where we show the $\varphi_2$-averaged data of the second and third columns of Fig.~\ref{fig:force2d} 
together with theoretical results from Eqs.~(\ref{eq:proj-f1-er}) and (\ref{eq:proj-f1-e1}) 
using $g_0(1)=\tfrac{\zeta_0}{2\pi}$ and $g_1(1)=\tfrac{\zeta_1}{\pi}$. 
In both Figs.~\ref{fig:force2d-averaged}(b) and \ref{fig:force2d-averaged}(c), we additionally show 
least-squares fits to the simulation data in accord with the respective special form of the 
theoretical expressions in Eqs.~(\ref{eq:proj-f1-er}) and (\ref{eq:proj-f1-e1}), 
i.e., $f_{\rm a}+f_{\rm b}\cos(\theta)$ in Fig.~\ref{fig:force2d-averaged}(b) and 
$f_{\rm a}+f_{\rm b}\cos(\theta)+f_{\rm c}\cos(2\theta)$ in Fig.~\ref{fig:force2d-averaged}(c). 
Interestingly, these fits show much better agreement with the simulations than the theoretical 
predictions based on the $\zeta_i$. This finding might hint at problems in identifying the $g_i(1)$ 
with the $\zeta_i$, which we did for the theoretical predictions in Fig.~\ref{fig:force2d-averaged}, 
although the pair interaction in the simulations is not completely steep. However, the 
observation confirms the general $\theta$ dependence of the projected conditional 
force just up to the second order. 
Note that the data shown in Fig.~\ref{fig:force2d-averaged}(c) are also shown 
in Fig.~\ref{fig:force1_e1}(a) along spherical cuts around the tagged particle. 

%+++++++++++++++++++++++++++++++++++++++
\begin{figure*}
  \includegraphics[width=16.0cm]{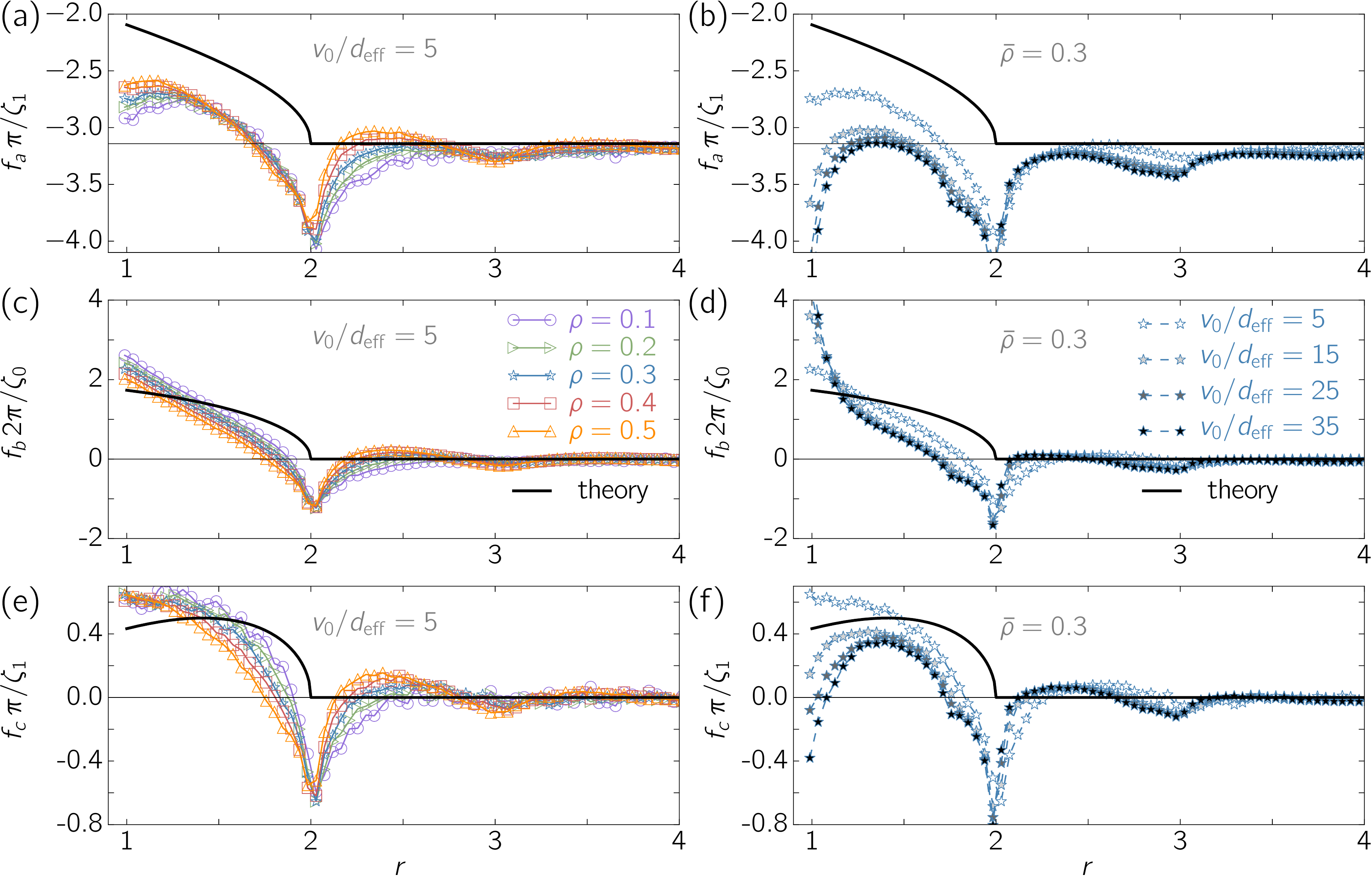}
	\caption{\label{fig:force1_e1_modes}
	Coefficients (a) and (b) $f_{\rm a}\pi/\zeta_1$, (c) and (d) $f_{\rm b}2\pi/\zeta_0$, and 
	(e) and (f) $f_{\rm c}\pi/\zeta_1$ from theory and simulations as defined in 
	Eq.~(\ref{eq:proj-f1-e1}). Theoretical curves are given by Eqs.~(\ref{eq:coef-fa})-(\ref{eq:coef-fc}) 
	and simulation results follow from least-squares fits as shown in Fig.~\ref{fig:force2d-averaged}. 
	Coefficients are shown at (a), (c), and (e) constant propulsion speed $v_0$ dependent on the 
	number density $\bar{\rho}$ and (b), (d), and (f) constant number density dependent on the propulsion speed. }
\end{figure*}
%+++++++++++++++++++++++++++++++++++++++

\begin{figure}
  \includegraphics[width=8.0cm]{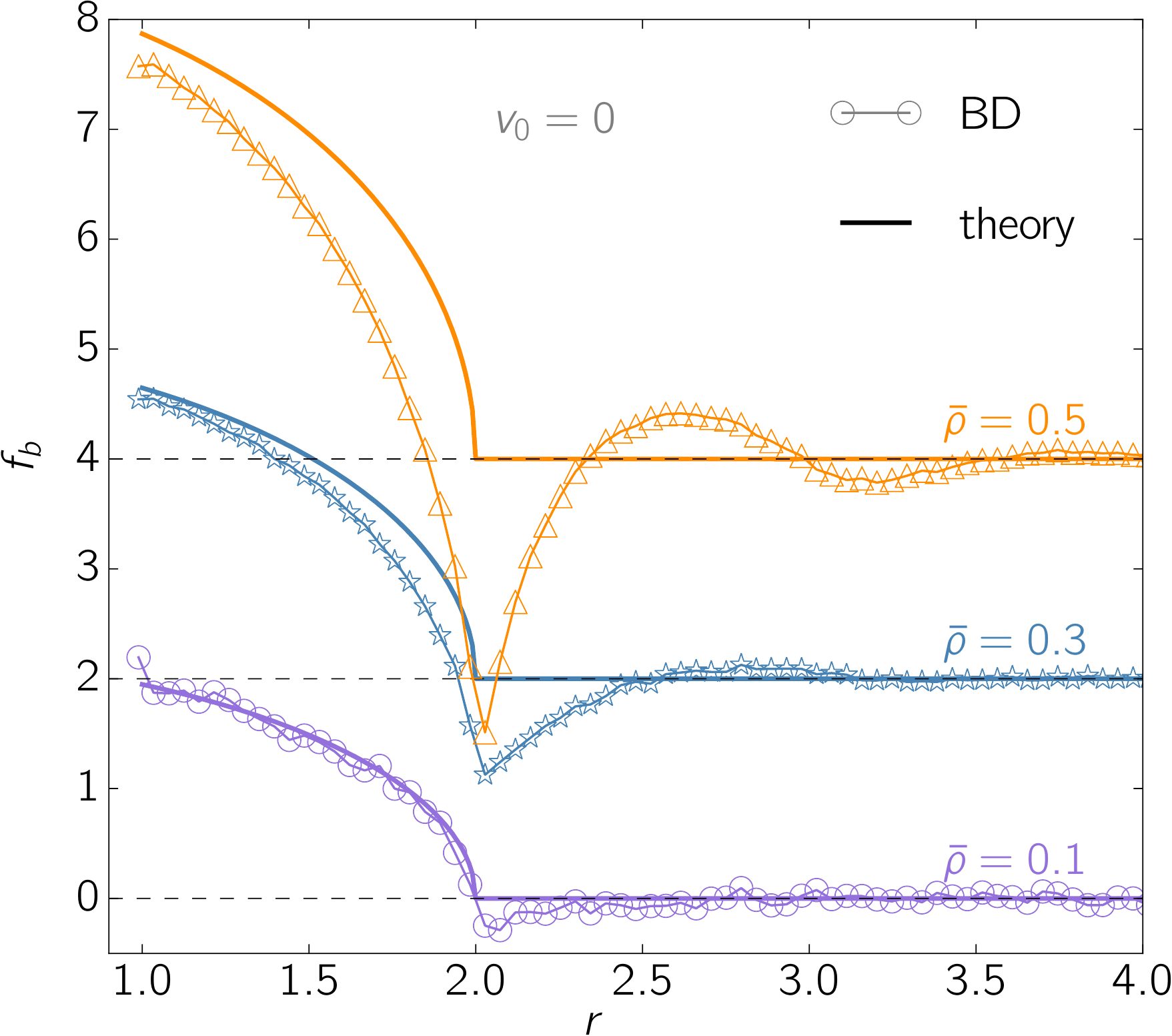}
	\caption{\label{fig:passive_disks} 
	Coefficient $f_{\rm b}$ from theory and simulations as defined in 
	Eq.~(\ref{eq:proj-f1-e1}) for a system of passive disks. Similar to Fig.~\ref{fig:force1_e1_modes}(c), 
	we show the coefficient $f_{\rm b}$ dependent on the number density $\bar{\rho}$. 
	Note that data are shifted to enhance readability. The dashed lines mark zero for 
	each number density $\bar{\rho}$. }
\end{figure}

According to the previously confirmed $\theta$ dependence of the conditional force, 
now we study the fitting of our simulation data with Eq.~(\ref{eq:proj-f1-e1}) in more detail. 
At the same time, we study the previously mentioned collapse of data onto uniform 
curves in Figs.~\ref{fig:force1_e1}(b)-\ref{fig:force1_e1}(d). 
For this purpose, we determine the fitting parameters 
$f_{\rm a}$, $f_{\rm b}$, and $f_{\rm c}$ from least-squares fits of 
Eq.~(\ref{eq:proj-f1-e1}) to the simulation data. We show the resulting parameters for certain 
combinations of propulsion speed $v_0$ and density $\bar{\rho}$ in Fig.~\ref{fig:force1_e1_modes} 
and for passive disks with $v_0=0$ in Fig.~\ref{fig:passive_disks}. As discussed previously, we 
replace the parameters $g_i(1)$ within our theory in Eqs.~(\ref{eq:coef-fa})-(\ref{eq:coef-fc}) by 
the parameters $\zeta_i$ via the relations in Eqs.~(\ref{eq:rel-zeta-g0}) and 
(\ref{eq:rel-zeta-gi}), because the $\zeta_i$ are more natural for our simulations of not 
completely hard disks. The resulting theoretical predictions of the coefficients are also shown 
in Figs.~\ref{fig:force1_e1_modes} and \ref{fig:passive_disks} and they read 
\begin{align}
	f_{\rm a}(r) &= \left\{ \begin{array}{lcl} \frac{\zeta_1}{\pi} \Big(\arccos\big(\frac{r}{2}\big)-\pi\Big) 
		& \quad & r\leq 2 , \\ -\zeta_1 & \quad & r>2 , \end{array} \right. \label{eq:coef-fa-2nd} \\
	f_{\rm b}(r) &= \left\{ \begin{array}{lcl} \frac{\zeta_0}{2\pi} \sqrt{4-r^2} 
		& \quad & r\leq 2 , \\ 0 & \quad & r>2 , \end{array} \right. \label{eq:coef-fb-2nd} \\
	f_{\rm c}(r) &= \left\{ \begin{array}{lcl} \frac{\zeta_1}{\pi} \sqrt{4-r^2} \,\frac{r}{4} 
		& \quad & r\leq 2 , \\ 0 & \quad & r>2 . \end{array} \right. \label{eq:coef-fc-2nd}
\end{align}
Note that in Fig.~\ref{fig:force1_e1_modes} we have removed the linear dependence 
of the coefficients $f_{\rm a}$ and $f_{\rm c}$ on $\zeta_1$ and of $f_{\rm b}$ on $\zeta_0$ 
by plotting $f_{\rm a}(r)\pi/\zeta_1$, $f_{\rm b}(r)2\pi/\zeta_0$, and $f_{\rm c}(r)\pi/\zeta_1$. 
In these cases, our theory in Eqs.~(\ref{eq:coef-fa-2nd})-(\ref{eq:coef-fc-2nd}) predicts 
a collapse of the data at different number densities $\bar{\rho}$ and propulsion 
speeds $v_0$ to unique and solely $r$-dependent curves, because the dependences on density and 
propulsion speed are only contained in the parameters $\zeta_i$. 

In accord with this prediction, we find the simulation data in Fig.~\ref{fig:force1_e1_modes} 
to be rather independent of the number density in Figs.~\ref{fig:force1_e1_modes}(a), 
\ref{fig:force1_e1_modes}(c), and \ref{fig:force1_e1_modes}(e). However, for different propulsion 
speeds the data in Figs.~\ref{fig:force1_e1_modes}(b), \ref{fig:force1_e1_modes}(d), and 
\ref{fig:force1_e1_modes}(f) show deviations from a collapse, especially at small propulsion speeds 
and small separation $r$. 
Also in contrast to our theory, the simulation data show detailed radial structure with a 
pronounced negative peak at $r\approx 2$. This peak matches with our 
observation of a dip in the data shown in Fig.~\ref{fig:force1_e1}, which we 
explained by an interaction between the tagged first particle and a second particle 
via intermediate third particles. The dip is not described in our theory because we closed the Smoluchowski 
equation using the assumption from Eq.~(\ref{eq:assumption-of-spherical-step-function}) that neglects 
higher-order structure between the second particle and third particles 
and we did not consider situations where two 
particles interact via more than one intermediate particle at all. For instance, we have shown a 
snapshot from our simulations in Fig.~\ref{fig:sketch-coordinates}(c), where two particles $1$ and $2$ 
interact via two additional particles $3$ and $4$. 

We have seen that at large separations $r$ the conditional force $\vec{F}_1$ takes the constant 
value $-\bar{\rho}\zeta_1\hat{e}_1$. Accordingly, theory and simulation show a value of $-\pi$ (thin line) 
in Figs.~\ref{fig:force1_e1_modes}(a) and \ref{fig:force1_e1_modes}(b). 
This negative value of the coefficient $f_{\rm a}$ 
describes a $\theta$-averaged effective slow down of the tagged first particle due to third particles, 
which scales with the propulsion speed $v_0$ via the parameter $\zeta_1$. The nonuniform shape of 
the projected force $\hat{e}_1\cdot\vec{F}_1$ with respect to the location $\theta$ of the second 
particle is captured in the higher-mode coefficients $f_{\rm b}$ and $f_{\rm c}$. Of course, the relative 
location of the second particle becomes irrelevant at large $r$, where both coefficients vanish as shown in 
Figs.~\ref{fig:force1_e1_modes}(c)-\ref{fig:force1_e1_modes}(f). 
At $r\approx 1$, the coefficient $f_{\rm b}$ reaches 
a maximum, which is related to an acceleration of the tagged particle if the second particle is ahead 
and to a slowdown if it is behind. We argued that the second particle blocks contributions from 
third particles from the respective direction. Interestingly, we find a change in sign for $f_{\rm b}$ 
at $r\approx 2$ in Figs.~\ref{fig:force1_e1_modes}(c) and \ref{fig:force1_e1_modes}(d). 
Accordingly, the tagged particle now is effectively accelerated by the third particles if the second 
particle is located behind and it is slowed down if the second particle is ahead. 
At small separations $r\approx 1$, the simulation data in Fig.~\ref{fig:force1_e1_modes}(b) 
show a strong deviation from the theoretical prediction that increases with increasing propulsion speed. 
The simulation shows a much stronger average deceleration of the tagged particle than predicted by the 
theory. At the same time, we also find a stronger anisotropy in Fig.~\ref{fig:force1_e1_modes}(d) at high 
propulsion speeds $v_0$ than predicted by our theory. In this situation of particle contact, third 
particles are more likely located in simultaneous contact with both tagged particles 1 and 2 than elsewhere, 
which follows from the Kirkwood closure in Eq.~(\ref{eq:kirkwood-gs}) together with the fact that pair 
distributions of (at least passive) hard disks have 
maxima at particle contact. Again, this situation is underestimated in our theory due to the 
assumption from Eq.~(\ref{eq:assumption-of-spherical-step-function}) such that third particles are less 
likely located in contact with both tagged particles in comparison to simulations. As a result, the tagged 
particle is predicted to be slowed down less by third particles in our theory 
if the second particle is located ahead at $(r,\theta)=(1,0)$, 
which we can observe in Figs.~\ref{fig:force1_e1_modes}(b), \ref{fig:force1_e1_modes}(d), and 
\ref{fig:force1_e1_modes}(f). Note that for the total slowdown of a particle, we have to 
sum up the contributions from all coefficients. The fact that even Fig.~\ref{fig:force1_e1_modes}(f) 
shows strong deviations from the theoretical curve might hint at a problem with cutting the expansion 
of $g(r,\theta)$ after the first mode in Sec.~\ref{sec:expansion-g}. 

%+++++++++++++++++++++++++++++++++++++++
%+++++++++++++++++++++++++++++++++++++++

The deviations between simulations and theory might further hint at problems that arise when the 
Kirkwood closure is applied to systems of active particles. For a comparison with the active systems, 
we plot the coefficient $f_{\rm b}$ for passive disks without self-propulsion in 
Fig.~\ref{fig:passive_disks}. The other coefficients $f_{\rm a}$ and $f_{\rm c}$ vanish for passive disks. 
Note that, in comparison 
to Fig.~\ref{fig:force1_e1_modes}, we do not divide $f_{\rm b}$ by $\zeta_0$ and, accordingly, the 
theoretical curves do not collapse to one unique curve. We furthermore have to use the $\zeta_0$ as 
an input for our theoretical curves, because we do not independently achieve the parameters from our theory. 
To improve visibility, we have shifted the data and marked the original zero by horizontal lines, 
respectively, for each number density. 
%%%%
In the limit $r=1$, we now observe good agreement between theory and simulation for all shown number 
densities. For increasing density, however, still a dip at $r\approx 2$ develops, but it is less 
pronounced in comparison to the one observed for self-propelled disks at higher propulsion speed. 
Of course, deviations between Figs.~\ref{fig:passive_disks} and \ref{fig:force1_e1_modes}(d) at 
$r\approx 1$ could also appear due to the fact that in Fig.~\ref{fig:force1_e1_modes}(d) the coefficient 
$f_{\rm b}$ is divided by the parameter $\zeta_0$, but such deviations should appear at all values of $r$, 
especially at higher ones where the simulation confirms the theory. 

In conclusion, we could identify mainly two effects that lead to the observed behavior of the coefficients 
$f_{\rm a}$, $f_{\rm b}$, and $f_{\rm c}$ at the positions $r\approx 1$ and $r\approx 2$, i.e., at particle contact 
and at the position of the discussed dip. The dip mainly originates from the three-body structure between 
both tagged particles 1 and 2 and a third particle. It does not appear in our theory, because we 
neglect the secondary structure beyond volume exclusion between particles 2 and 3 by our approximation from 
Eq.~(\ref{eq:assumption-of-spherical-step-function}). The behavior at $r\approx 1$ is described well 
for passive disks within our theory. For self-propelled disks the deviations from the theory originate, 
next to the missing contribution of structure between particles 2 and 3, from stopping 
the expansion of $g(r,\theta)$ in Sec.~\ref{sec:expansion-g} at a certain order and from applying the 
Kirkwood closure in active systems.

\subsection{Pair-distribution function}
\label{sec:pair-distribution-function}

In our simulations we have full access to the pair-distribution function $g(r,\theta)$ and, 
using Eqs.~(\ref{eq:zeta0}) and (\ref{eq:zeta1}), to the parameters $\zeta_i$ of its expansion 
in Eq.~(\ref{eq:g-expansion}). 
In the preceding section we used these parameters from our simulations to test our analytic 
theory. The theory is derived from the Smoluchowski equation (\ref{eq:eom5c1}), which can 
also be solved numerically without applying the 
simplified closure discussed in Sec.~\ref{sec:simplification-hard-disks}, which 
we applied to obtain analytical results. When we use Eq.~(\ref{eq:eom5c1}) in order 
to obtain data for $g(r,\theta)$, the conditional forces $\vec{F}_i$ 
that enter Eq.~(\ref{eq:eom5c1}) are given in Eqs.~(\ref{eq:force1-C2}) and (\ref{eq:force2-C2}) 
for our system of self-propelled hard disks. 

We solve Eq.~(\ref{eq:eom5c1}) using a forward-time and center-space scheme \cite{press_book_1992} 
on a numerical grid with $(r_i,\theta_{ij})\in[1,R]\times[0,2\pi]$. For the radial $r$ component we 
use $N_r=600$ equidistant grid points and set $R=6$. For the angular $\theta$ component we use equally 
distributed $N_{\theta,i}$ grid points at each radial index $i$, respectively, such that the spacing 
$r_i(\theta_{i,j+1}-\theta_{i,j})$ between two points of indices $j$ and $j+1$ is smaller than or equal 
to $\Delta_{\rm num}=0.1$, i.e., we set $N_{\theta,i}=\lceil 2\pi r_i/\Delta_{\rm num} \rceil$. 
Here $\lceil a \rceil$ denotes the rounded up integer of $a$. Since the number of grid points 
$N_{\theta,i}$ in the angular direction depends on the radial index $i$, we use linear interpolation along 
the angular $\theta$ coordinate to perform the center-space scheme in the radial direction. 
At the boundaries with $r=1$ and $r=R$, we use Neumann boundary conditions, i.e., 
we apply the no-flux condition which is given in Eq.~(\ref{eq:no-flux}) for $r=1$. 
Outside the grid, we assume $g(r,\theta;t)=0$ when $r<1$ and 
$g(r,\theta;t)=1$ when $r>R$. As an initial configuration at time $t_0$, we have chosen 
$g(r_i,\theta_j;t_0)=1$ for $r\geq 1$. We then run $N_t=3\times 10^5$ time steps of size $dt=10^{-5}$ to 
achieve a final variation of 
$\lVert \partial_t g(r_i,\theta_j;t_{N_t}) \rVert \lesssim 0.02$, 
where $\lVert a_{ij} \rVert$ denotes the maximum norm of $a_{ij}$. 
We call this final state the steady-state solution of Eq.~(\ref{eq:eom5c1}). 

%+++++++++++++++++++++++++++++++++++++++
\begin{figure*}
	\centering
  \includegraphics[width=16.0cm]{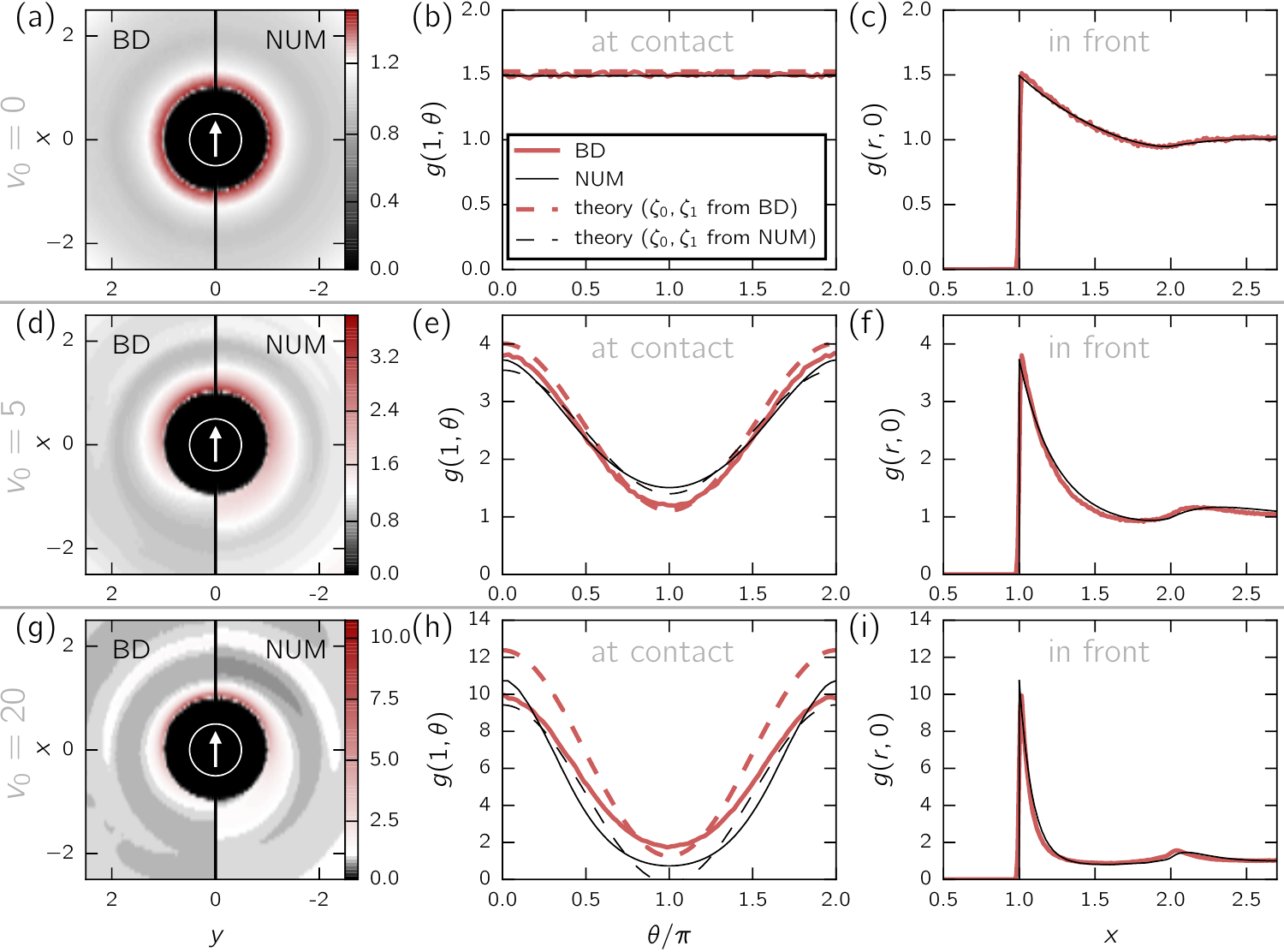}
	\caption{\label{fig:full-g}
	Pair-distribution functions $g(r,\theta)$ around a self-propelled particle that is located at $(0,0)$ 
	and swims in the positive $x$ direction. We show data at number density $\bar{\rho}=0.3$ and 
	at propulsion speeds (a)-(c) $v_0=0$, (d)-(f) $v_0=5$, and (g)-(i) $v_0=20$. 
	The full function is shown in (a), (d), and (g) and has the symmetry $g(r,\theta)=g(r,-\theta)$. 
	Accordingly, we show data obtained from our BD simulations (BD) on the left half of the plot and numerical 
	results from our theory (NUM) on the right half of the same plot. 
	(b), (e), and (f) Values at particle contact, where we use the effective diameter 
	as the particle-particle separation in our simulations. In addition, we show the theoretical predictions 
	according to the parameters $\zeta_0$ and $\zeta_1$, which we have calculated from the respective 
	data. Values are presented in Table~\ref{tab:zeta-values}. 
	(c), (f), and (i) Function $g(x,0)$ along the positive $x$ axis in front of the tagged particle. 
	}
\end{figure*}
%+++++++++++++++++++++++++++++++++++++++

We show our numerical results in comparison to results from our BD simulations in 
Fig.~\ref{fig:full-g} for three propulsion speeds 
$v_0=0$ [Figs.~\ref{fig:full-g}(a)-\ref{fig:full-g}(c)], 
$v_0=5$ [Figs.~\ref{fig:full-g}(d)-\ref{fig:full-g}(f)], and 
$v_0=20$ [Figs.~\ref{fig:full-g}(g)-\ref{fig:full-g}(i)]. 
Again, we chose the same density $\bar{\rho}=0.3$ as studied 
previously in Fig.~\ref{fig:force1_e1_modes}. The steady-state pair-distribution 
function $g(r,\theta)$ is symmetric in the angle $\theta$ and for this reason we draw half planes 
only for our BD data (left) and numerical data (right) in Figs.~\ref{fig:full-g}(a), 
\ref{fig:full-g}(d), and \ref{fig:full-g}(g). The plots 
in these panels are parametrized 
by $(x,y)=(r\cos\theta,r\sin\theta)$, where the respective length unit is the (effective) 
particle diameter $d_{\rm eff}$ for the BD data and $\sigma$ for the numerical data. 
Furthermore, we show data at particle contact in Figs.~\ref{fig:full-g}(b), \ref{fig:full-g}(e), 
and \ref{fig:full-g}(h), i.e., along the line with $r=1$ in Figs.~\ref{fig:full-g}(a), 
\ref{fig:full-g}(d), and \ref{fig:full-g}(g), and in front of the particle in Figs.~\ref{fig:full-g}(c), 
\ref{fig:full-g}(f), and \ref{fig:full-g}(i), i.e., along the positive $x$ axis in 
Figs.~\ref{fig:full-g}(a), \ref{fig:full-g}(d), and \ref{fig:full-g}(g). At finite propulsion speed, 
the data, especially at particle contact, 
show a peak in the pair-distribution function ahead of the tagged particle and a depletion behind it. 
While the numerical solutions for $g(r,\theta)$ are overall converged, the exact depth of the 
minimum at $g(1,\pi)$ in this depletion area is still sensitive with respect to the grid discretization. 
For the employed grid, the solutions fit well with the results from the numerical simulations. 
Small deviations between both solutions from theory and simulations are visible for the 
finite propulsion speeds $v_0=5$ and $v_0=20$ in Figs.~\ref{fig:full-g}(d)-\ref{fig:full-g}(i), 
especially behind the tagged particle.

\begin{table}
\center
\caption{\label{tab:zeta-values}
Values for $\zeta_0$ and $\zeta_1$, extracted from our BD simulations and from our numerical 
solutions of Eq.~(\ref{eq:eom5c1}) as shown in Fig.~\ref{fig:full-g}. 
Values are rounded to two digits after the decimal. }
\begin{tabular}{ l|cc|cc }
\toprule[1.5pt]
~~~~~~~~~ & ~~~$\bar{\rho}$~~~ & ~~~$v_0$~~~ & ~~~$\zeta_0$~~~ & ~~~$\zeta_1$~~~ \\
\hline
BD simulations  & $0.3$ & $0$  & $9.60$  & -- \\
~   & ~     & $5$  & $16.08$ & $4.55$ \\
~   & ~     & $20$ & $42.88$ & $17.46$ \\
\hline
Numerical solutions & $0.3$ & $0$  & $9.38$  & -- \\
~   & ~     & $5$  & $15.54$ & $3.37$ \\
~   & ~     & $20$ & $29.08$ & $15.08$ \\
\midrule
\bottomrule[1.5pt]
\end{tabular}
\end{table}

Having at hand data for the full pair-distribution function $g(r,\theta)$, we can calculate the 
corresponding parameters $\zeta_0$ and $\zeta_1$ using Eqs.~(\ref{eq:zeta0}) and (\ref{eq:zeta1}). 
For our BD and numerical results from Fig.~\ref{fig:full-g} we present these parameters in 
Table~\ref{tab:zeta-values}. We observe that the numerical results from our theory underestimate 
both the mean-value parameter $\zeta_0$ and the anisotropy parameter $\zeta_1$. The gap between 
the BD and numerical data increases with increasing propulsion speed $v_0$. Furthermore, we can 
use the expansion from Eq.~(\ref{eq:g-expansion}) to determine the pair-distribution function 
$g(r=1,\theta)$ at particle contact from the parameters $\zeta_0$ and $\zeta_1$ via 
the relations $\zeta_0=2\pi g_0(1)$ and $\zeta_1=\pi g_1(1)$ for hard disks from 
Eqs.~(\ref{eq:rel-zeta-g0}) and (\ref{eq:rel-zeta-gi}). 
We show these theoretical curves for parameters $\zeta_i$ obtained from the BD and numerical data 
in Figs.~\ref{fig:full-g}(b), \ref{fig:full-g}(e), and \ref{fig:full-g}(f) by dashed lines. 
While we observe only minor deviations in Figs.~\ref{fig:full-g}(b) and \ref{fig:full-g}(e) between 
the theoretical lines and the corresponding simulation and numerical data, respectively, we find 
strong deviations in Fig.~\ref{fig:full-g}(h). Here the 
theoretical curve fed by the parameters from BD predicts much higher values in front of the tagged 
particle than BD itself. Even the average value that is related to $\zeta_0$ is higher than that found 
in the simulation data. This finding originates from the not completely hard pair potential in 
Eq.~(\ref{eq:wca-potential}) that we used in our simulations, for which the relations between 
the $\zeta_i$ and the $g_i(1)$ do not hold strictly such that the relation becomes inaccurate at 
high propulsion speed. Indeed, the resulting 
parameters $g_i(1)$ describe the effective hard pair distribution at contact, which is higher 
than the smeared-out distribution of the ``softer'' interaction in our simulations. Of course, the 
identities hold for hard disks and, accordingly, the theoretical curve fed by the parameters 
from the numerical data is obeyed on average. However, the predicted values at $\theta=\pi$ behind the tagged 
particle take slightly negative values, because the shape of the line at particle contact cannot be captured 
completely by the first two modes of the expansion of the pair-distribution function. The latter 
shows a very wide minimum for the numerical data in Fig.~\ref{fig:full-g}(h) in comparison to the BD data. 
Apart from the deviations between the numerical and BD results behind the tagged particle, both agree well 
for the distribution of second particles ahead of the tagged particle, as shown in 
Figs.~\ref{fig:full-g}(c), \ref{fig:full-g}(f), and \ref{fig:full-g}(i). 
In agreement with our simulations, the numerically obtained pair-distribution function even 
shows maxima at positions ahead of the self-propelled particle with $r\approx 2$, $r\approx 3$, $r\approx 4$, 
and so on (only the first is shown in Fig.~\ref{fig:full-g}), as we expect from 
the discussion of the structure of the conditional force $\vec{F}_1$ along with 
Fig.~\ref{fig:force1_e1} and in Sec.~\ref{sec:test-theory}. At high propulsion speeds, 
these maxima are located slightly farther away from the tagged particle in the numerical results 
when compared to the BD results, as we find in Fig.~\ref{fig:full-g}(c).

%%%%%%%%%%%%%%%%%%%%%%%%%%%%%%%%%%%%%%%%%%%%%%%%%%%%%%%%%%%%%%%%%%%%%%%%%%%%%%%%%%%%%%%%%%%%%%%%%%
%%%%%%%%%%%%%%%%%%%%%%%%%%%%%%%%%%%%%%%%%%%%%%%%%%%%%%%%%%%%%%%%%%%%%%%%%%%%%%%%%%%%%%%%%%%%%%%%%%

\section{Discussion}

In the previous sections we have derived an analytical theory for the microscopic structure of 
active Brownian particles (ABPs) that interact via hard pair potentials. We have analyzed this theory 
by testing its predictions using BD simulations and by solving the underlying Smoluchowski equation 
(\ref{eq:eom5c1}) numerically. In this context, we have studied pair-distribution functions and 
conditional three-body forces between the active particles. 

In Sec.~\ref{sec:simplification-hard-disks} we have identified two main contributions to the averaged 
conditional force $\vec{F}_1$ that acts on a tagged first particle in the presence of a second tagged 
particle from all remaining particles. The corresponding two main directions of the conditional force 
$\vec{F}_1$ are the direction of the (normalized) separation vector $\hat{e}_r=\vec{r}/|\vec{r}\,|$ between 
both tagged particles and the direction of self-propulsion $\hat{e}_1$ of the first particle. 
We discussed that these directions originate from the excluded volume due to the presence of the 
second particle and from the directed motion of the first particle and the resulting collisions with 
surrounding particles. 
From another perspective, both directions further correspond to the splitting of the total force that 
acts on a tagged particle into the conditional force $\vec{F}_1$ 
and the contribution $\vec{F}_{12}$ from the second particle. 
In our study we found the dependence of $\vec{F}_{12}$ on the angular position $\theta$ at small 
propulsion speeds of the same order as that of $\vec{F}_1$, but 
we found the force $\vec{F}_{12}$ and its anisotropy almost independent of 
the propulsion speed $v_0$. In contrast, we observed a strong dependence on 
the propulsion speed for the anisotropy of $\hat{e}_r\cdot\vec{F}_1$. 
This might lead to situations where, at sufficiently high propulsion speeds, 
the free energy can be reduced by clustering of particles with a second particle ahead. 

Indeed, anisotropic correlations due to the self-propulsion of the particles are a key 
ingredient for the motility-induced phase separation. However, the system of ABPs still 
is described solely by the scalar fields number density and propulsion speed. For this reason, 
the system is still classified as scalar active matter \cite{tiribocchi_prl115_2015}, 
as pointed out in the Introduction. 

In our analysis of simulation results in Sec.~\ref{sec:bd-sim}, we found the orientation of 
the second tagged particle to be rather unimportant. If it is taken into account, the dependence 
on this orientation is strongest for the collision term of the conditional force $\vec{F}_1$ along 
the propulsion direction $\hat{e}_1$ at particle-particle contact ($r\approx 1$) and for the 
excluded-volume term along $\vec{r}$ at intermediate particle separations of $r\approx 1.5$. 
However, the relative position of the second tagged particle in comparison to the first one is very 
important, i.e., the angle $\theta$ at which the second particle is located around the first one and the 
separation $r$. While the anisotropic angular shape of the conditional force $\vec{F}_1$ at 
a given separation $r$ overall is described well by only two or three modes within our theory at 
small and large separations, the theory does not describe additional modes that appear at 
intermediate separations of $r\approx 1.5$. Thus, it might be necessary to also take higher modes 
into account when intermediate particle separations dominate. 

The radial dependence of the conditional force $\vec{F}_1$ on the separation $r$ 
is most interesting ahead of the tagged first particle in its direction of self propulsion. 
In Fig.~\ref{fig:force1_e1} we have shown that this force has a dip that develops at 
$r\gtrsim 2$ when the propulsion speed $v_0$ of the particles is increased. Our theory does not 
predict this dip, as we have shown in Figs.~\ref{fig:force1_e1_modes} and \ref{fig:passive_disks}. 
As discussed previously, the dip develops at a separation $r\gtrsim 2$ of the first and the second 
particle, where a third particle fits in between them. This intermediate particle leads 
to a strong repulsive force between the two tagged particles. 
The latter is not described by our theory, because we neglect structural 
correlations between second and third particles beyond volume exclusion for the 
calculation of the conditional forces $\vec{F}_i$ by applying the approximation in 
Eq.~(\ref{eq:assumption-of-spherical-step-function}). 
For example, Fig.~\ref{fig:passive_disks} shows the coefficient $f_{\rm b}$ in a system of passive 
disks that, according to Eqs.~(\ref{eq:proj-f1-er}) and (\ref{eq:coef-fb}), corresponds to the negative 
strength of the conditional force onto the first particle. When the second particle is located close 
to the first particle, it blocks third particles from interacting with the first particle in a 
certain area, as shown in Fig.~\ref{fig:sketch-2particles}. The amount of surface that is blocked for 
third particles is described by the angle $\theta^*$. This angle decreases when the separation $r$ 
increases until the separation between the first and second particles becomes $r\geq 2$. 
Our theory does not assume a higher probability of finding third particles in the 
vicinity of the second particle due to our assumption made in 
Eq.~(\ref{eq:assumption-of-spherical-step-function}) such that the conditional force $\vec{F}_1$ 
vanishes for all $r\geq 2$. In the simulations, the probability of finding third particles in the vicinity of the 
second particle is higher than average and for this reason the simulation data show a pronounced dip 
around $r=2$ in Figs.~\ref{fig:force1_e1_modes} and \ref{fig:passive_disks}. 
Note that problems also arise when the assumption from 
Eq.~(\ref{eq:assumption-of-spherical-step-function}) is used to solve 
Eq.~(\ref{eq:eom5c1}) self-consistently, because the angle $\theta^*$ that enters the theory is not 
continuously differentiable.

Our theory is based on the two-body Smoluchowski equation, which we closed on the three-body 
level using Kirkwood's approximation. We tested this theoretical framework by calculating its 
steady-state solutions numerically without applying the assumption from 
Eq.~(\ref{eq:assumption-of-spherical-step-function}). The numerical 
solution of our theory does predict the anisotropic structure in the pair-distribution function 
at moderate propulsion speed that originates from the dip in the conditional forces, 
as we have shown in Sec.~\ref{sec:pair-distribution-function} and in Fig.~\ref{fig:full-g}. 
At higher propulsion speeds the agreement between our theoretical approach and our simulation 
results becomes poorer as we observed in Table~\ref{tab:zeta-values} for the $\zeta_i$ and in 
Fig.~\ref{fig:full-g}. 
Furthermore, our theoretical approach cannot capture situations where two particles interact via 
more than one intermediate particle, because the approach is based on a three-body closure. 
It has been shown that such multiparticle interactions are relevant 
for active systems \cite{hanke_pre88_2013,chou_pre91_2015} such that it might be necessary to 
extend our theory to the four-body level. In our simulations we have observed situations 
where at least four particles interact, as exemplified in the snapshot in 
Fig.~\ref{fig:sketch-coordinates}(c).

Our theory in Eqs.~(\ref{eq:coef-fa-2nd})-(\ref{eq:coef-fc-2nd}) successfully 
predicts a collapse of the data at different number densities $\bar{\rho}$ and sufficiently 
high propulsion speeds $v_0$ onto unique and solely $r$-dependent curves. The data of 
all involved coefficients $f_{\rm a}/\zeta_1$, $f_{\rm b}/\zeta_0$, and $f_{\rm c}/\zeta_1$ 
collapse onto a unique curve, respectively, because the dependences on density and propulsion 
speed are only contained in the parameters $\zeta_i$. The observed collapse in the simulation 
data in Figs.~\ref{fig:force1_e1_modes} and \ref{fig:passive_disks} confirms this dependence 
of the $\zeta_i$ on the propulsion speed and the number density and thus indicates the role 
of the $\zeta_i$ as order parameters in systems of ABPs. In Fig.~\ref{fig:force1_e1_modes} 
we found the strongest deviations from the collapse to a unique curve at small propulsion 
speeds and particle separations $r$. Via a comparison between active and passive disks, 
we argued that these deviations arise from problems with the applied Kirkwood closure in 
combination with the activity of the ABPs. 

We have seen that our theory predicts the general shape of the anisotropic conditional force 
and of the two-body distribution function. Its analytic form is based on the expansion of the 
pair-distribution function in Eq.~(\ref{eq:g-expansion}) and on the involved parameters $g_i(r)$. 
For the hard-disk potential from Eq.~(\ref{eq:hard-disk-potential}), we found the 
equalities in Eqs.~(\ref{eq:rel-zeta-g0}) and (\ref{eq:rel-zeta-gi}) 
between the contact values $g_i(1)$ of the expansion coefficients and 
the parameters $\zeta_i$, which are defined in Eqs.~(\ref{eq:zeta0}) and (\ref{eq:zeta1}). 
The latter can also be obtained directly from not perfectly hard potentials 
like the WCA potential in Eq.~(\ref{eq:wca-potential}), which we have used in 
our BD simulations with a very strong coefficient $\epsilon=100 k_{\rm B}T$ to simulate a system of 
effectively hard disks. Note that for almost hard interactions gradients become steep and 
the numerical resolution must be chosen appropriately. 
In Sec.~\ref{sec:pair-distribution-function} we have seen for our 
simulations of not completely hard disks that the equalities between the parameters $\zeta_i$ and 
values of $g(r)$ at contact do not hold, because contact is not well defined for soft potentials. 
Indeed, the $\zeta_i$ are defined as pair distributions weighted with the derivative of the pair 
potential and for this reason they represent effective hard-disk coefficients that correspond to 
the coefficients in the expansion of $g(r)$ at contact for hard disks. 
In our theory we neglected all higher modes in the 
expansion in Eq.~(\ref{eq:g-expansion}) such that all $\zeta_i$ for $i>1$ vanish. 
In fact, we have shown that the first two modes $\zeta_0$ and $\zeta_1$ in a system of 
ABPs already predict the main directions of the acting forces, describe the general anisotropic 
shape of the conditional forces, and explain the effective swimming speed. 
The linear relation between the parameter $\zeta_1$ and the propulsion speed 
$v_0$ can be seen either from the projection of the conditional force 
$\hat{e}_1\cdot\vec{F}_1$, in Figs.~\ref{fig:force1_e1}(b)-\ref{fig:force1_e1}(d), where all data 
collapse when it is divided by the propulsion speed $v_0$, or from the fact that 
the projected conditional force approaches $-\bar{\rho}\zeta_1$ at large separations $r>2$. 
This finding further agrees with previous work \cite{bialke_epl103_2013,stenhammar_prl111_2013}, 
where $\zeta_1$ is discussed to be proportional to $v_0$ with a 
proportionality factor of approximately one. 

Our analytic theory does not independently predict the parameters $\zeta_i$. 
However, we have shown that our analytic theory is predictive for given $\zeta_i$ and that 
results are in good agreement with our simulations. 
We further found that numerical solutions of Eq.~(\ref{eq:eom5c1}) together with the conditional 
forces from Eqs.~(\ref{eq:force1-C2}) and (\ref{eq:force2-C2}) agree well with our simulation 
data. The corresponding parameters $\zeta_0$ and $\zeta_1$ fit those obtained from 
our simulations up to moderate propulsion speeds of $v_0\approx 5$, but we find strong deviations at 
larger $v_0\approx 20$. As possible reasons we discussed closing the Smoluchowski equation on the 
three-body level, the Kirkwood approximation, neglecting additional structure between second and third 
particles, and stopping the expansion of $g(r)$ after the second-order term. 
In any case, obtaining the parameters $\zeta_i$ from our numerical solutions would make the 
theory independent such that it could be used to predict MIPS or the pressure in active systems 
from the combined knowledge of the pair interactions, the free propulsion 
speed, and the density without any additional input.

\section{Conclusion}

In this work we have studied two-body and especially three-body correlations and conditional 
forces in systems of active Brownian particles. Based on the many-body Smoluchowski equation, 
we have developed a theoretical framework that we closed on the three-body level. Applied to 
the special case of hard-particle interactions, we have derived analytical expressions for 
conditional three-body forces and identified preferred directions of these forces with respect to 
the direction of propulsion of tagged particles. We have verified our theory in 
a detailed comparison with Brownian dynamics computer simulations, for which we have reported 
three-body forces for the first time. 
In this context we also have discussed discrepancies between the modeling of active particles with hard 
pair-interaction potentials and soft or almost hard potentials. 
As a consequence, theoretical models for active systems that are based on hard 
interaction potentials must be handled carefully when they are applied to systems of not 
completely hard particles. 
For future work it might be interesting to also study effective interaction potentials within 
our theory as performed in recent work \cite{farage_pre91_2015}. 

We further have identified the range of validity and limitations of our theory. 
While we have found generally good agreement between theory and simulations at sufficient 
small propulsion speeds, we have observed qualitative and quantitative deviations 
that increase with the strength of the propulsion speed. 
We have discussed these deviations to be caused most probably by 
(i) the Kirkwood closure which we have applied in our theory, 
(ii) neglecting higher modes in an expansion of the pair-distribution function, and 
(iii) an assumption where we effectively neglect correlations beyond volume exclusion 
between a second particle and its surrounding ones. 
For this reason, future work should study how to improve closures and test the influence of higher modes. 
Note that improving on closures could also mean closing the Smoluchowski equation on an even higher 
level than we have done. 

We have shown that our theory captures
many effects that occur in systems of Brownian swimmers. 
Based on only the first two modes $\zeta_0$ and $\zeta_1$ in the expansion of the pair-distribution function, 
our analytic theory already successfully predicts the main directions of the conditional three-body forces, 
their linear dependence on the propulsion speed, and the effective swimming speed. 
These findings are in agreement with previous work. 
However, our approach does not yield independent expressions for $\zeta_0$ and $\zeta_1$. 
Such expressions would be necessary to obtain {\it a priori} theoretical predictions without further input 
of correlations. In any way, our theory has at least two levels of approximation. The first level is 
more general and is reached after closing our theory in Sec.~\ref{sec:closure-twobody-level} and 
applying it to the special case of hard disks in Sec.~\ref{sec:special-case-hard-disks}. The 
second level is reached by applying the additional approximation from 
Eq.~(\ref{eq:assumption-of-spherical-step-function}) in Sec.~\ref{sec:simplification-hard-disks}, 
which allows us to derive analytical expressions for the conditional three-body forces. 
We have shown that a numerical solution of our theory already on the first level is in very good 
agreement with our simulations such that the necessary parameters $\zeta_i$ in general could be 
obtained from numerical calculations. 

In a next step, the parameters $\zeta_i$ could be 
used to predict physical quantities, for instance, 
phase separations like MIPS \cite{cates_arcmp6_2015,speck_epjst225_2016} 
and the pressure in active Brownian systems \cite{solon_prl114_2015,bialke_prl115_2015,speck_pre93_2016}. 
Another step could be the transfer of our findings to self-propelled Brownian swimmers in three dimensions. 
In conclusion, our detailed study of correlations in suspensions of active repulsive disks is a step 
towards an emerging liquid-state theory of scalar active matter.

%%%%%%%%%%%%%%%%%%%%%%%%%%%%%%%%%%%%%%%%%%%%%%%%%%%%%%%%%%%%%%%%%%%%%%%%%%%%%%%%%%%%%%%%%%%%%%
%%%%%%%%%%%%%%%%%%%%%%%%%%%%%%%%%%%%%%%%%%%%%%%%%%%%%%%%%%%%%%%%%%%%%%%%%%%%%%%%%%%%%%%%%%%%%%

\section*{Acknowledgements}  

We acknowledge financial support from the German Research Foundation 
within the priority program SPP 1726 (Grant No. SP 1382/3-2) 
and through Grant No. INST 39/963-1 FUGG. 
We thank R.~Wittkowski for helpful discussions. 
Further, we gratefully acknowledge computing time granted on the supercomputer 
Mogon at Johannes Gutenberg University Mainz and 
acknowledge support from the state of Baden-W\"urttemberg through bwHPC.

%%%%%%%%%%%%%%%%%%%%%%%%%%%%%%%%%%%%%%%%%%%%%%%%%%%%%%%%%%%%%%%%%%%%%%%%%%%%%%%%%%%%%%%%%%%%%%
%%%%%%%%%%%%%%%%%%%%%%%%%%%%%%%%%%%%%%%%%%%%%%%%%%%%%%%%%%%%%%%%%%%%%%%%%%%%%%%%%%%%%%%%%%%%%%
%%%%%%%%%%%%%%%%%%%%%%%%%%%%%%%%%%%%%%%%%%%%%%%%%%%%%%%%%%%%%%%%%%%%%%%%%%%%%%%%%%%%%%%%%%%%%%

\appendix

\section*{Appendix: Mode expansion}

In Eq.~(\ref{eq:g-expansion}) we expand the pair-distribution function $g(r,\theta)$ 
from Eq.~(\ref{eq:definition-g}) in Fourier modes. 
Accordingly, we find $\langle g(r,\theta) \rangle_{\theta} = g_0(r)$ and 
the projections of the conditional force $\vec{F}_1$ from Eq.~(\ref{eq:force1fb}) 
onto the directions $\hat{e}_r$ and $\hat{e}_\theta$ become 
\begin{align}
	\hat{e}_r\cdot\vec{F}_1(r,\theta) =& 
	-\bar{\rho} \sum_{k=0}^{\infty} g_k(\sigma) A^{\rm c}_k(r) \cos(k\theta) , 
	\label{eq:proj-force1-to-r} \\
%%%%%%%%%%%%%%%
	\hat{e}_\theta\cdot\vec{F}_1(r,\theta) =& 
		-\bar{\rho} \sum_{k=0}^{\infty} g_k(\sigma) A^{\rm s}_k(r) \sin(k\theta)  
	\label{eq:proj-force1-to-theta} . 
\end{align}
The mode-expansion coefficients $A^{\rm c}_k$ and $A^{\rm s}_k$ are defined using the 
$r$-dependent angle $\theta^*$ from Eq.~(\ref{eq:excluded-volume-angle}) by 
\begin{align}
A^{\rm c}_k(r) \cos(k\theta) &= 
	\int_{\theta^*}^{2\pi-\theta^*} d\varphi \cos(\varphi) \cos\big(k(\varphi+\theta)\big) 
	\label{eq:mode-expansion-factor-cos} , \\
%%%%%
A^{\rm s}_k(r) \sin(k\theta) &= 
	\int_{\theta^*}^{2\pi-\theta^*} d\varphi \sin(\varphi) \cos\big(k(\varphi+\theta)\big) 
	\label{eq:mode-expansion-factor-sin} .
\end{align}
The integrals in Eqs.~(\ref{eq:mode-expansion-factor-cos}) and (\ref{eq:mode-expansion-factor-sin}) 
can be performed analytically and, for $k\in\{0,1\}$ and $1\leq r\leq 2$, result in 
\begin{align}
	A^{\rm c}_0(r) &= -2\sin(\theta^*) , \\
	A^{\rm c}_1(r) &= \Big(\pi-\theta^*-\sin(\theta^*)\cos(\theta^*)\Big) , \\
%	A^{\rm c}_k(r) &= 2\frac{\cos(k\theta^*)\sin(\theta^*)
%		-k\sin(k\theta^*)\cos(\theta^*)}{k^2-1} , \\
%%%
	A^{\rm s}_0(r) &= 0 , \\
	A^{\rm s}_1(r) &= -\Big(\pi-\theta^*+\sin(\theta^*)\cos(\theta^*)\Big) . %, \\
%	A^{\rm s}_k(r) &= 2\frac{\sin(k\theta^*)\cos(\theta^*)-
%		k\cos(k\theta^*)\sin(\theta^*)}{k^2-1}  
\end{align}
In general, for $r>2$ all coefficients vanish except for 
$A^{\rm c}_1=\pi$ and $A^{\rm s}_1=-\pi$. 
At particle-particle contact with $r=1$ the first coefficients are 
$A^{\rm c}_0(\sigma)=-\sqrt{3}$, 
$A^{\rm c}_1(\sigma)=\tfrac{2\pi}{3}-\tfrac{\sqrt{3}}{4}$, and 
$A^{\rm s}_1(\sigma)=-\tfrac{2\pi}{3}-\tfrac{\sqrt{3}}{4}$.

When we insert the full expansion of the pair-distribution function $g(r,\theta)$ 
from Eq.~(\ref{eq:g-expansion}) into the no-flux condition from Eq.~(\ref{eq:no-flux}) 
we achieve a set of equations, one for each occurring Fourier component $\cos(k\theta)$. 
Solving the equation for $k=0$ with respect to $g_1(1)$, we obtain 

\begin{align}
g_1(1) &= \frac{1}{K\bar{\rho}} \bigg( 
   v_0 \pm \sqrt{v_0^2 + 8 K\bar{\rho} J
		 }
		  \bigg) \label{eq:no-flux-res1} , \\
 J &= 
	 %\Big(
	 -\frac{\bar{\rho}}{4}\sum_{k=2}^{\infty}g_k(1)g_k(1) A^{\rm c}_k 
  	 + \sqrt{3}g_0(1)g_0(1)\bar{\rho}+ g_0'(1) %\Big) 
		 , 
\end{align}
where $g_0'(1)=\tfrac{\partial}{\partial r}g_0(r)|_{r=1}$ and 
$K=(8\pi-3\sqrt{3})/6\approx 3.323$. 
In the limit of vanishing propulsion speed $v_0\to 0$ all $g_k$ for $k>0$ must vanish. 
Accordingly, $J$ must vanish and solely the plus sign in front of the square root in 
Eq.~(\ref{eq:no-flux-res1}) holds. 
%Furthermore, we find $g_0'(1)=-\sqrt{3}g_1(1)g_1(1)\bar{\rho}$ in this limit.
%
A rearrangement of Eq.~(\ref{eq:no-flux-res1}) and using $\zeta_1=\pi g_1(1)$ with $v_0>0$ leads to 
\begin{align}
v_0 \Big(1 - \bar{\rho}\frac{\zeta_1}{v_0}\frac{K}{\pi} \Big) &= 
   \mp \sqrt{v_0^2 + 8 K\bar{\rho} J
		 } . \label{eq:no-flux-res1b}
\end{align}
The form of Eq.~(\ref{eq:no-flux-res1b}) is interesting for the effective propulsion 
speed in the context of MIPS, as discussed by Bialk\'e {\it et al.} \cite{bialke_epl103_2013} and by Stenhammar 
{\it et al.} (above Fig.~2 in their work) \cite{stenhammar_prl111_2013}.

%%%%%%%%%%%%%%%%%%%%%%%%%%%%%%%%%%%%%%%%%%%%%%%%%%%%%%%%%%%%%%%%%%%%%%%%%%%%%%%%%%%%%%%%%%%%%%
%%%%%%%%%%%%%%%%%%%%%%%%%%%%%%%%%%%%%%%%%%%%%%%%%%%%%%%%%%%%%%%%%%%%%%%%%%%%%%%%%%%%%%%%%%%%%%

%\bibliography{./literature}

%Control: production of eprint (-1) disabled
%

\end{document}